\documentclass [12pt, a4paper]{article}

\usepackage[left=1.0in,top=1.5in,right=1.0in,bottom=1.5in]{geometry}
\usepackage[utf8]{inputenc}
\usepackage{dsfont}
\usepackage{amsmath,amssymb}
\usepackage{amsthm}
\usepackage{bm}
\usepackage{color}
\usepackage{graphics}
\usepackage{graphicx}
\usepackage{booktabs}
\usepackage{ascmac}
\usepackage{pb-diagram}
\usepackage{amscd} 
\usepackage{natbib}
\usepackage{enumerate}
\usepackage{authblk}
\usepackage{here}
\usepackage{multirow}
\usepackage{diagbox}
\usepackage{hyperref}
\hypersetup{
    unicode=true,
    colorlinks=true,
    linkcolor=red,
    citecolor=blue,
    filecolor=magenta,      
    urlcolor=cyan
    }

\newcommand{\lp}{\left(}  
\newcommand{\rp}{\right)} 
\newcommand{\llp}{\left\{}  
\newcommand{\rrp}{\right\}}  
\newcommand{\lllp}{\left[}  
\newcommand{\rrrp}{\right]} 

\newtheorem{theorem}{Theorem}
\newtheorem{example}{Example}

\newtheorem{corollary}{Corollary}

\newtheorem{lemma}{Lemma}

\title{Identification enhanced generalised linear model estimation with nonignorable missing outcomes}

\author[1]{Kenji Beppu}
\author[2]{Jinung Choi}
\author[1]{Kosuke Morikawa}
\author[2]{Jongho Im\footnote{Corresponding author: ijh38@yonsei.ac.kr}}

\affil[1]{Graduate School of Engineering Science, Osaka University, Osaka, Japan}
\affil[2]{Department of Statistics and Data Science, Yonsei University, Seoul, Republic of Korea}

\date{}

\begin{document}
\maketitle
\thispagestyle{empty}

\begin{abstract}
Missing data often result in undesirable bias and loss of efficiency. These issues become substantial when the response mechanism is nonignorable, meaning that the response model depends on unobserved variables. To manage nonignorable nonresponse, it is necessary to estimate the joint distribution of unobserved variables and response indicators. However, model misspecification and identification issues can prevent robust estimates, even with careful estimation of the target joint distribution.
In this study, we modeled the distribution of the observed parts and derived sufficient conditions for model identifiability, assuming a logistic regression model as the response mechanism and generalized linear models as the main outcome model of interest.
{More importantly, the derived sufficient conditions do not require any instrumental variables, which are often assumed to guarantee model identifiability but cannot be practically determined beforehand. To analyze missing data in applications, we propose practical guidelines and sensitivity analysis to determine the response mechanism.}
Furthermore, we present the performance of the proposed estimators in numerical studies and apply the proposed method to two sets of real data: exit polls from the 19th South Korean election and public data collected from the Korean Survey of Household Finances and Living Conditions.
\end{abstract}

%==Section 1==%
\section{Introduction}

Handling missing data is an important issue in many research areas because inappropriate analysis of missing data may yield erroneous results.
For example, ``The Strengthening Analytical Thinking for Observational Studies Initiative'', a large-scale collaborative research project involving over 100 experts from diverse fields of biostatistics research, identified nine crucial topics in observational studies \citep{sauerbrei2014strengthening}; missing data problems are one of the nine critical topics. Handling missing data can become more problematic when the response mechanism is nonignorable or missing not at random (MNAR), in which missingness is caused by the value that would have been observed \citep{ibrahim2005, kim2011semiparametric, wang2014instrumental,little2019statistical}. 

The nonignorable response mechanism is frequently encountered in practical applications and has been extensively studied in various fields such as RNA-sequencing data analysis \citep{hicks2018missing}, school psychology research \citep{baraldi2010introduction}, cost-effectiveness analysis \citep{leurent2018sensitivity}, counselling psychology research \citep{parent2013handling}, clinical trials \citep{hazewinkel2022sensitivity}, and information systems research \citep{peng2023handling}.
For example, \cite{peng2023handling} reported how information systems researchers deal with missing values and presented six scenarios with R\&D data from the compustat database and online reviews in the field of information systems.

Nonignorable nonresponse is extremely challenging to analyse because it often requires modelling the response model as well as the distribution of the sampled data to further construct the observed likelihood. Moreover, model identifiability may be violated even for a simple model. To guarantee identifiability, several studies assumed the existence of a covariate, referred to as a non-response instrumental variable or a shadow variable \citep{wang2014instrumental,zhao2015semiparametric,  miao2016varieties, zhao2018optimal,li2021efficient,shetty2021avoid}. Although the instrumental variable is beneficial in several cases, selecting it from observed data is an elusive task. In contrast, \cite{miao2016identifiability} and \cite{cui2017identifiability} derived certain conditions to identify models without using instrumental variables. These studies assumed parametric models for the distribution of the outcome variable for the complete data and for the response mechanism and subsequently derived sufficient conditions for model identifiability.
However, this assumption is subjective, and its validity cannot be verified using only observed data. 
Mitigating this assumption is crucial in applied research.

Instead of modelling the distribution of complete data, numerous recently developed methods have modelled the distribution of observed data, referred to as the respondents' outcome model \citep{kim2011semiparametric, riddles2016propensity, morikawa2021semiparametric,li2021efficient,shetty2021avoid}.
This modelling is advantageous because the observed data are available; consequently, we can select a better model for the candidates by using information criteria based on the observed data, such as the Akaike information criterion (AIC), Bayesian information criterion (BIC), or other variable selection methods, such as the adaptive lasso \citep{zou2006adaptive}.
The application of Bayes' theorem yields an explicit expression of the nonrespondents' outcome model from the assumed respondents' outcome model and response mechanism. 
This alternative expression for the joint distribution of the outcome variable and response indicator is often called Tukey's representation \citep{franks2020}. 

Employing Tukey's representation and \cite{im2017multiple}'s imputation approach, this study derives the nonrespondents’ outcome model in the form of a generalised linear model without any instrumental or shadow variables when the observed data can be fitted as a generalised linear model and the response mechanism follows a logistic distribution. To estimate the model parameters, we employ fractional imputation (FI), which is among the most beneficial tools in missing data analysis for solving estimating equations \citep{kim2011parametric, im2018fhdi}. Detailed FI estimation procedures are introduced along with variance estimation by applying the results of \cite{riddles2016propensity}.
Given these points, deriving identification conditions without relying on the existence of instrumental variables is useful and accessible to all fields of research.

While our method is an efficient tool for identifying model parameters, it faces the challenge of confirming that the response model strictly adheres to a logistic form, even though its functional structure can be validated. 
This indicates that our parametric model approach may be susceptible to model misspecification \citep{little1985note, kenward1998selection}. To address this concern, we conduct a thorough investigation through several simulation studies. In the simulation studies, we compare our method to proxy pattern mixture models (PPM) \citep{andridge2011proxy, andridge2015using, little2020measures}, which are known for their robustness against response model misspecification and utilize sensitivity analysis based on the pattern-mixture model \citep{little1994class}. Our investigation confirms that our method is robust for the outcome model and performs effectively, provided there is no significant misspecification in the response model.

{One approach to handling missing data, as exemplified by the PPM, avoids identifying parameters that distinguish ignorability and non-ignorability.
However, this approach often requires extrapolation to assign values to these parameters, which introduces its own set of challenges.
In contrast, our proposed method assumes a parametric structure for the model, which allows us to estimate the parameters without the need for extrapolation.
Comparing these approaches is crucial, as the proposed method proves effective when the use of parametric models is supported by domain knowledge.}

The remainder of this paper is organised as follows: In Section \ref{basic setup}, we derive the sufficient conditions to identify single-outcome models and subsequently extend them to mixture models.
The observed likelihood and estimation procedure using FI are presented in Section \ref{Estimation}. Numerical examples including several simulation studies and real data applications, are presented in Section \ref{Simulation} and \ref{Real Data}, respectively. Section \ref{Discussion} summarises the concluding remarks. All technical proofs are presented in the Supplementary Materials.

%==Section 2==%
\section{Identification condition}\label{basic setup}
\subsection{Basic setup}
Let \((\bm{x}_i, y_i, \delta_i)\) for \(i = 1, \ldots, n\) be \(n\) independently realized values of the random variables \((\bm{x}, y, \delta)\), where \(y\) denotes a response variable subject to missingness, \(\delta\) denotes a response indicator for \(y\) that takes the value 1 if \(y\) is observed and 0 if it is missing, and \(\bm{x}\) represents a vector of fully observed covariates. We assume that the response mechanism follows a logistic model:
\begin{align}
\text{logit}\left\{P(\delta=1 \mid \bm{x}, y; \bm{\alpha}, \beta)\right\} = h(\bm{x}; \bm{\alpha}) + \beta y, \label{Resp}
\end{align}
where \(\text{logit}(z) := \log\left(z/(1-z)\right)\) for all \(0 < z < 1\), \(h(\bm{x}; \bm{\alpha})\) is injective with respect to \(\bm{\alpha}\), and is known up to a finite-dimensional parameter \(\bm{\alpha}\). In nonignorable missing data analysis, several previous studies have employed this logistic model \citep{kim2011semiparametric, shao2016semiparametric, 2105.12921}, and {\cite{hirano1998combining} investigates the identification of this type of response model in the context of panel data with the presence of refreshment samples.} 

Suppose that the outcome variable distribution of the respondent, given the covariates $[y_i\mid \bm{x}_i, \delta_i=1]$ belongs to the exponential family in the form
\begin{align}
    f(y_i\mid \bm{x}_i,\delta_i=1;\bm{\gamma})=\exp\lllp \tau\{y_i\theta_i -b(\theta_i)\}+c(y_i;\tau)  \rrrp, \label{expo}
\end{align}
   where $\theta_i=\theta(\eta_i),\ \eta_i=\sum_{l=1}^{L} \eta_l(\bm{x}_i)\kappa_l,\ \bm{\kappa}=(\kappa_1,\ldots,\kappa_L)^{\top},$ and $\ \bm{\gamma}=(\tau,\bm{\kappa}^{\top})^{\top}$. 
   This class includes several distributions such as binomial, normal, gamma, and Poisson. Function $\theta$ is defined according to the purpose of the statistical analysis.

When the outcome model of the respondents belongs to the exponential family in (\ref{expo}) and the response mechanism follows the logistic model in (\ref{Resp}), the outcome model of the non-respondents also belongs to the same exponential family, but with a different parameterization: 
\begin{align}
    f(y \mid \bm{x},\delta=0 )&=f(y \mid \bm{x},\delta=1 )\frac{\exp\llp-h(\bm{x};\bm{\alpha})-\beta y\rrp}{\int \exp\llp-h(\bm{x};\bm{\alpha})-\beta y\rrp f(y \mid \bm{x},\delta=1 )dy} \notag\\
    &\propto \exp\lllp \tau \llp y\lp \theta  - \beta \tau^{-1} \rp -b\lp \theta  - \beta \tau^{-1} \rp\rrp +c(y;\tau) \rrrp.\label{f0 calculation} 
\end{align}

{We say that the model is identifiable if
  \begin{align*}
     P(\delta=1\mid \bm{x},y;\bm{\alpha},\beta)f(y\mid \bm{x};\bm{\alpha},\beta,\bm{\gamma})
     =P(\delta=1\mid \bm{x},y;\bm{\alpha}^{\prime},\beta^{\prime})f(y\mid \bm{x};\bm{\alpha}^{\prime},\beta^{\prime},\bm{\gamma}^{\prime}) \ \  \mathrm{w.p.1}
  \end{align*}
implies $(\bm{\alpha}^{\top},\beta,\bm{\gamma}^{\top})^{\top}=({\bm{\alpha}^{\prime}}^{\top},{\beta^{\prime}},{\bm{\gamma}^{\prime}}^{\top})^{\top}$.
Simple parametric models of $P(\delta=1 \mid y,\bm{x})$ and $f(y\mid \bm{x})$ cannot be identified, e.g., Example $1$ in \cite{wang2014instrumental}.
In the following subsections, we present the identification conditions with examples of commonly used outcome models.}

%In the case that the model is identifiable, the parameters $(\bm{\alpha}^{\top},\beta,\bm{\gamma}^{\top})^{\top}$ are estimable by solving the mean score equation $\bar{S}(\bm{\alpha},\beta)=0$, where
%\begin{align*}
%    \bar{S}(\bm{\alpha},\beta)&=\sum_{i=1}^{n} \lllp \delta_iS_1(\bm{\alpha},\beta;\bm{x}_i,y_i)+(1-\delta_i) E\llp S_0(\bm{\alpha},\beta;\bm{x}_i,Y)\mid \bm{x}_i,\delta_i=0  \rrp \rrrp,\\
%    S_{\delta}(\bm{\alpha},\beta;\bm{x},y)&=\frac{\delta-P(\delta=1\mid \bm{x},y)}{P(\delta=1\mid \bm{x},y)\llp 1-P(\delta=1\mid \bm{x},y) \rrp} \frac{\partial P(\delta=1\mid \bm{x},y;\bm{\alpha},\beta)}{\partial (\bm{\alpha}^{\top},\beta)^{\top}},
%\end{align*} 
%and $\hat{\bm{\gamma}}=\mathrm{arg}\max_{\bm{\gamma}}\sum_{i=1}^{n} \delta_i\log  f(y_i\mid \bm{x}_i,\delta_i=1;\bm{\gamma})$.

\subsection{Single outcome model} \label{single outcome}
In this section, we derive sufficient conditions representing model identifiability, considering the outcome models that belong to the exponential family (\ref{expo}), and extend the result to its mixture.
The following theorem is a general result of model identifiability for the outcome model (\ref{expo}).

\begin{theorem} \label{identify}
Suppose that the response mechanism is (\ref{Resp}) and the distribution of $[y\mid \bm{x},\delta=1]$ is identifiable with a density that belongs to the exponential family (\ref{expo}).
Then, this model is identifiable {if and only if} the following condition holds for all $\bm{\alpha},\bm{\alpha}^{\prime},\beta,\beta^{\prime},\bm{\gamma}$:
\begin{align}
\varphi(\bm{\alpha},\beta,\bm{\gamma})=\varphi(\bm{\alpha}^{\prime},\beta^{\prime},\bm{\gamma}) \Rightarrow  \beta=\beta^{\prime},  \label{condition}
\end{align}
where
\begin{align*}
   \varphi(\bm{\alpha},\beta,\bm{\gamma})=  h(\bm{x};\bm{\alpha})-\tau b\llp \theta\lp \sum_{l=1}^{L} \eta_l(\bm{x})\kappa_l\rp -\frac{\beta}{\tau} \rrp .
\end{align*}
\end{theorem}
A vital implication of Theorem \ref{identify} is that the identifiability of the model is equivalent to that of $\varphi(\bm{\alpha},\beta,\bm{\gamma})$. Furthermore, the model identification of $\varphi$ can be verified only with respect to $\beta$.
Based on Theorem \ref{identify}, we can conveniently check the identification conditions for almost all distributions belonging to the exponential family even if the covariates $\bm{x}$ contain both discrete and continuous variables.
When the covariates $\bm{x}$ contain only discrete variables, we can determine whether the number of unknown variables $(\bm{\alpha}^{\top},\beta)^{\top}$ is less than or equal to the number of values taken by the covariates $\bm{x}$.
Additionally, we provide Corollary \ref{normal identification}, which specifically assumes that the outcome model follows a normal distribution because it requires careful attention, as detailed in Example \ref{normal ex}.  

\begin{corollary}\label{normal identification}
Suppose that the response mechanism is (\ref{Resp}), $h(\bm{x};\bm{\alpha})$ is a polynomial, and the outcome model of respondent is $ N(\mu(\bm{x};\bm{\gamma}),\sigma^2)$, where the link function represents the identity $\theta(\eta)=\eta$ such that $\mu(\bm{x};\bm{\gamma})=\sum_{l=1}^{L} \eta_l(\bm{x})\kappa_l$. Then, condition (\ref{condition}) holds if an index $l=1,\ldots,L$ exists such that $\eta_l(\bm{x})$ is continuous and not represented by $h(\bm{x};\bm{\alpha})$ for all $\bm{\alpha}$. 
\end{corollary}

For better understanding of Theorem \ref{identify} and Corollary \ref{normal identification}, we introduce some examples that are commonly used in the generalised linear model (GLM).

\begin{example} \label{normal ex}
Considering the same setting as in Corollary \ref{normal identification}, the function $\varphi$ in Theorem \ref{identify} can be expressed as 
\begin{align*}
\varphi(\bm{\alpha},\beta,\bm{\gamma})=h(\bm{x};\bm{\alpha})+\beta\sum_{l=1}^{L} \eta_l(\bm{x})\kappa_l-\frac{\sigma^2\beta^2}{2}.
\end{align*}
Condition (\ref{condition}) holds if $\sum_{l=1}^{L} \eta_l(\bm{x})\kappa_l$ contains a term not included in $h(\bm{x};\bm{\alpha})$.
For instance, $\sum_{l=1}^{L} \eta_l(\bm{x})\kappa_l=\kappa_0+\kappa_1x+\kappa_2x^2\ (\kappa_2\neq0)$ and $h(\bm{x};\bm{\alpha})=\alpha_0+\alpha_1x$ satisfy this condition if the covariate $x$ is continuous. If covariate $x$ is binary, the identification condition does not hold because of the three unknown variables, $(\alpha_0,\alpha_1,\beta)$.
In addition, the model is not identifiable for $\kappa_2=0$ even if covariate $x$ is continuous.
This is identical to the example in \cite{morikawa2021semiparametric}. 
\end{example}

\begin{example} \label{binomial Example}
Suppose $[y\mid \bm{x},\delta=1]\sim B(1,p(\bm{x}))$, which belongs to the exponential family, with $\tau=1,\ \theta=\log p/(1-p),\ b(\theta)=\log\{1+\exp(\theta)\},\ c(y;\tau)=0,\ \theta(\eta)=\eta$. Accordingly, we check the identification of this model, and function $\varphi$ in Theorem \ref{identify} can be expressed as
\begin{align*}
\varphi(\bm{\alpha},\beta,\bm{\gamma})=h(\bm{x};\bm{\alpha})-\log\llp 1+\exp\lp -\beta+\sum_{l=1}^{L} \eta_l(\bm{x})\kappa_l\rp \rrp.
\end{align*}
For example, condition (\ref{condition}) holds if the polynomials $h(\bm{x};\bm{\alpha})$ and $\eta_l(\bm{x})$ contain continuous variables. 

For discrete nonmeasurement variables such as sex and area, the outcome models for each nonmeasurement variable should be assumed, as discussed in Section \ref{Real Data}. For instance, if $z$ denotes sex and $x$ represents another continuous covariate, we can model various mean structures: $\kappa_{01}+\kappa_{11}x$ for males and $\kappa_{02}+\kappa_{12}x$ for females rather than $\kappa_0+\kappa_1x+\kappa_2z$. In these cases, we have sufficient conditions for model identifiability. 
\end{example}

\begin{example} \label{categorical normal} 
Suppose that nonmeasurement categorical variables occur in $D$ cases with $z\ (=1,2,\ldots, D)$ indicating one of the $D$ cases, the response mechanism is (\ref{Resp}) and $h(\bm{x};\bm{\alpha})=\sum_{d=1}^{D}h_d(\bm{x};\bm{\alpha}_d)$, and the outcome density of the respondents can be expressed as
\begin{align*}
    \prod_{d=1}^{D} \lllp \frac{1}{\sqrt{2\pi \sigma_d^2}} \exp\llp  -\frac{\lp y-\mu_d(\bm{x};\bm{\kappa}_d) \rp^2}{2\sigma_d^2} \rrp \rrrp^{I(z=d)},  
\end{align*}
where $\bm{\kappa}_d$ and $\sigma_d^2$ represent the mean function and variance parameter, respectively, and $I(\cdot)$ denotes the indicator function. As it is normal distribution with mean $\sum_{d=1}^{D} I(z=d)\mu_d(\bm{x};\bm{\kappa}_d)$ and variance $\sum_{d=1}^{D} I(z=d)\sigma_d^2$, according to Example \ref{normal ex}, the function $\varphi$ in Theorem \ref{identify} can be stated as
\begin{align*}
\varphi(\bm{\alpha},\beta,\bm{\gamma})=\sum_{d=1}^{D}h_d(\bm{x};\bm{\alpha}_d)+\beta I(z=d)\mu_d(\bm{x};\bm{\kappa}_d)
-\frac{\beta^2}{2}I(z=d)\sigma_d^2.
\end{align*}
Similar to Corollary \ref{normal identification}, this model is identifiable if an index $l=1,\ldots,D$ exists, such that $ \mu_l(\bm{x};\bm{\kappa}_l) $ is continuous and not represented by $h_l(\bm{x};\bm{\alpha}_l)$ for all $\bm{\alpha}_l$.
\end{example}

\begin{example} \label{categorical logistic}
Suppose that nonmeasurement categorical variables occur in $D$ cases with $z\ (=1,2,\ldots,D)$ indicating one of the $D$ cases, the response mechanism is (\ref{Resp}) and $h(\bm{x};\bm{\alpha})=\sum_{d=1}^{D}h_d(\bm{x};\bm{\alpha}_d)$, and $[y\mid \bm{x},\delta=1]\sim B(1,p(\bm{x}))$, where
$
    \mathrm{logit}\{p(\bm{x})\}=\sum_{d=1}^{D} I(z=d)\sum_{l=1}^{L_d} \eta_{ld}(\bm{x})\kappa_{ld}.
$
The function $\varphi$ in Theorem \ref{identify} can be expressed as
\begin{align*}
\varphi(\bm{\alpha},\beta,\bm{\gamma})=\sum_{d=1}^{D}h_d(\bm{x};\bm{\alpha}_d)-\log\llp 1+\exp\lp -\beta+ \sum_{d=1}^{D} I(z=d)\sum_{l=1}^{L_d} \eta_{ld}(\bm{x})\kappa_{ld}   \rp \rrp.
\end{align*}
Similar to Example \ref{categorical normal}, the model can be identified more easily than in  the case without categorical measurement variables. 
\end{example}

%=Section 3.2=%
\subsection{Mixture outcome models} \label{mixture iden}
In this subsection, we derive sufficient conditions for model identifiability when the response mechanism is (\ref{Resp}), and the outcome model of the respondents $[y\mid \bm{x},\delta=1;\bm{\gamma}]$ is a mixture distribution of the exponential family (\ref{expo})
\begin{align}
     \sum_{k=1}^{K}\pi_k\exp\lllp \tau_k \llp y\theta_k -b(\theta_k)\rrp +c(y;\tau_k)  \rrrp , \label{mixture form}
\end{align}
where $\bm{\pi}=(\pi_1,\ldots,\pi_K)^{\top}$ represents the mixing proportion of the mixture models, i.e., $\sum_{i=k}^{K}\pi_k=1$ and $\pi_k\geq0$, $\theta_k=\theta(\eta_k)$ and $\bm{\tau}=(\tau_1,\ldots,\tau_K)^{\top}$ denote the model parameters, $\eta_k=\sum_{l=0}^{m(k)} \eta_{lk}(\bm{x})\kappa_{lk}$, $\bm{\kappa}_k=(\kappa_{0k},\kappa_{1k},
\ldots,\kappa_{m(k) k})^{\top}$ and $\bm{\kappa}=(\bm{\kappa}_1^{\top},\ldots,\bm{\kappa}_K^{\top})^{\top}$ are link functions and their parameters, $\bm{\gamma}=(\kappa^{\top},\tau^{\top},\pi^{\top},K)^{\top}$ is a vector of all of the parameters, and $m(k)+1$ indicates a dimension of the vector $\bm{\kappa}_k$. The following theorem is the most general result representing the identifiability of the mixture model, and its results are consistent with those of Theorem \ref{identify} for $K=1$.

\begin{theorem}\label{identify multi}
Suppose that the response mechanism is (\ref{Resp}) and the distribution of $[y\mid \bm{x},\delta=1]$ is (\ref{mixture form}) and identifiable.
Then, this model is identifiable {if and only if} the following condition holds for all $\bm{\alpha},\bm{\alpha}^{\prime},\beta,\beta^{\prime},\bm{\gamma}$:
\begin{align*}
    g(\bm{\alpha},\beta,\bm{\gamma})=g(\bm{\alpha}^{\prime},\beta^{\prime},\bm{\gamma}) \Rightarrow  \beta=\beta^{\prime},
\end{align*}
where
\begin{align*}
  g(\bm{\alpha},\beta,\bm{\gamma})=h(\bm{x};\bm{\alpha}) - \log\lllp \sum_{k=1}^{K} \pi_k\exp\llp -\tau_kb(\theta_k)+\tau_kb\lp \theta_k-\frac{\beta}{\tau_k} \rp \rrp \rrrp.
\end{align*}
\end{theorem}

Because the most popular and commonly used mixture model is a normal mixture, we discuss it in more detail the identification conditions for the case where $[y\mid \bm{x},\delta=1;\bm{\gamma}]$ follows a normal mixture distribution:
\begin{align}
    &[y\mid \bm{x}, \delta=1;\bm{\gamma}]\sim \sum_{k=1}^{K}\pi_k N(\mu_k(\bm{x};\bm{\kappa}_k),\sigma_k^2),  \label{4} 
\end{align}
where $\mu_k(\bm{x};\bm{\kappa}_k)$ denotes a polynomial $\sum_{l=0}^{m(k)} \eta_{lk}(\bm{x})\kappa_{lk}$, $\bm{\sigma}^2=(\sigma_1^2,\ldots,\sigma_K^2)^{\top}$ represents a vector of variance, and $\bm{\gamma}=(\bm{\kappa}^{\top},{\bm{\sigma}^2}^{\top},\bm{\pi}^{\top},K)^{\top}$ denotes a vector of all of the parameters. In this case, we obtain Corollary \ref{normal_mixture_cor} by applying Theorem \ref{identify multi}:

\begin{corollary} \label{normal_mixture_cor}
Suppose that the response mechanism is (\ref{Resp}) and the distribution of $[y\mid \bm{x},\delta=1]$ is (\ref{4}) and identifiable.
Then, this model is identifiable {if and only if} the following condition holds for all $\bm{\alpha},\bm{\alpha}^{\prime},\beta,\beta^{\prime},\bm{\gamma}$:
\begin{align*}
g(\bm{\alpha},\beta,\bm{\gamma})=g(\bm{\alpha}^{\prime},\beta^{\prime},\bm{\gamma}) \Rightarrow  \beta=\beta^{\prime},
\end{align*}
where
\begin{align*}
g(\bm{\alpha},\beta,\bm{\gamma})=h(\bm{x};\bm{\alpha}) - \log\llp \sum_{k=1}^{K} \pi_k\exp\lp -\beta \sum_{l=0}^{m(k)} \eta_{lk}(\bm{x})\kappa_{lk} +\frac{\beta^2\sigma_k^2}{2} \rp \rrp.
\end{align*}
\end{corollary}

{Hereafter, we consider a practically useful setup, where  $h(x;\bm{\alpha})=\sum_{j=0}^{J-1} \alpha_{j} h_{j}(\bm{x})$ and $\mu_k(x;\bm{\kappa}_k)=\sum_{l=0}^{m(k)} \eta_{lk}(\bm{x})\kappa_{lk}$, where $J$ is the dimension of $\bm{\alpha}$, the basis functions of $h_{j}(\bm{x})$ and $\eta_{lk}(\bm{x})$ have the form of polynomial function $\prod_{i=1 }^{\mathrm{dim}(\bm{x})} x_i^{s_i}$, $\mathrm{dim}(\bm{x})$ is a dimensional $\bm{x}$, and $s_i$ represents any nonnegative integer.
We define two classes of basis functions: 
\begin{align*}
    \mathcal{H}:&=\llp h_{j}(\bm{x})\,\bigg|\, j=0,1,\ldots, J-1 \rrp\cup\{ 1 \},\\
    \mathcal{M}:&=\llp \eta_{lk}(\bm{x}),\bigg|\, l=0,1,\ldots, m(k), \ k=1,\ldots,K  \rrp\setminus \mathcal{H},
\end{align*}
and decompose $\mu_k(\bm{x};\bm{\kappa}_k)$ into $\mu_k(\bm{x};\bm{\kappa}_k)=\mu_k^{\mathcal{H}}(\bm{x};\bm{\kappa}_k)+\mu_k^{\mathcal{M}}(\bm{x};\bm{\kappa}_k)$, where each $\mu_k^{\mathcal{H}}$ and $\mu_k^{\mathcal{M}}$ are constant multiple of the elements of $\mathcal{H}$ and $\mathcal{M}$, respectively.
For example, in the case of $h(x;\bm{\alpha})=2+4x$, $\mu_1(x;\bm{\kappa}_1)=3x^2$, and $\mu_2(x;\bm{\kappa}_2)=x+4x^3$, the definitions of the notation imply $\mathcal{H}=\{1,x\}$, $\mathcal{M}=\{ x^2, x^3 \}$, $\mu_1^{\mathcal{H}}=0$, $\mu_1^{\mathcal{M}}=3x^2$, $\mu_2^{\mathcal{H}}=x$, and $\mu_2^{\mathcal{M}}=4x^3$.}

When the distribution of $\bm{x}$ is discrete, comparing the number of unknown variables and the taken values of $\bm{x}$ is sufficient. Therefore, we consider only the continuous case:
\begin{itemize}
    \item[] $(\mathrm{C1})$ The distribution of $\bm{x}$ is continuous. 
\end{itemize}
The next theorem provides more rigorous conditions and identifiability results given the above setting.
\begin{theorem}\label{multi Logistic}
Suppose that the response mechanism is (\ref{Resp}) and the distribution of $[y\mid \bm{x},\delta=1]$ is identifiable and has a normal mixture density in (\ref{4}). Furthermore, we define three additional conditions:
\begin{itemize}
   \item[] $(\mathrm{C2})$ $\mathcal{M}\neq\emptyset$;
    \item[] $(\mathrm{C3})$ The sign of $\beta$ is known;
    \item[] $(\mathrm{C4})$ $\llp \mu_i^{\mathcal{M}}(\bm{x};\bm{\kappa}_i);\ i=1,\ldots,K\rrp\neq \llp -\mu_i^{\mathcal{M}}(\bm{x};\bm{\kappa}_i);\ i=1,\ldots,K\rrp $.
\end{itemize}
Then, this model is identifiable if $(\mathrm{C1})$--$(\mathrm{C2})$ and one of $(\mathrm{C3})$--$(\mathrm{C4})$ hold.
\end{theorem}

In application, confirming (C2) may be sufficient because violation of (C4) is rare in practical applications.

The following example shows an unidentifiable model that satisfies the condition $(\mathrm{C2})$, but does not satisfy $(\mathrm{C3})$--$(\mathrm{C4})$.

\begin{example}\label{double square gen}
Suppose the outcome model for respondents is $ \pi_1N(x^2,\sigma_1^2)+\pi_2N(-x^2,\sigma_2^2)$. Consider the following two response models:
\begin{align*}
    &\mathrm{Model \ 1}:\  \mathrm{logit} \llp P(\delta=1\mid x,y)\rrp =x+y;\\
    &\mathrm{Model \ 2}:\  \mathrm{logit} \llp P(\delta=1\mid x,y)\rrp =x-y.
\end{align*}
This model satisfies condition $(\mathrm{C2})$ because $\mathcal{M}=\{ x^2 \}$ and $\mathcal{H}=\{1, x \}$, but does not satisfy $(\mathrm{C3})$-$(\mathrm{C4})$ because we do not know the sign of $\beta$ and $\{ \mu_i^{\mathcal{M}}(\bm{x};\bm{\kappa}_i);\ i=1,\ldots,K\}=\{ x^2,-x^2 \}$.
The sufficient condition in Corollary \ref{normal_mixture_cor} is not satisfied if
\begin{align*}
   \frac{1}{2}\sigma_1^2+\log\pi_1=\frac{1}{2}\sigma_2^2+\log\pi_2 
\end{align*}
holds; thus, this model is unidentifiable.
\end{example}

The following example shows an unidentifiable model that does not satisfy (C2).

\begin{example}\label{single gen}
Suppose that the outcome model of the respondents is $\pi_1N(x,\sigma_1^2)+\pi_2N(2x,\sigma_2^2)$, considering two response models:
\begin{align*}
    &\mathrm{Model \ 1}:\ \mathrm{logit} \llp P(\delta=1\mid x,y)\rrp=x+y;\\
    &\mathrm{Model \ 2}:\ \mathrm{logit} \llp P(\delta=1\mid x,y)\rrp=4x-y.
\end{align*}
This model does not satisfy the condition $(\mathrm{C2})$ because $\mathcal{M}=\emptyset$ and $\mathcal{H}=\{1, x \}$.
The sufficient condition in Corollary \ref{normal_mixture_cor} is not satisfied if
\begin{align*}
   \frac{1}{2}\sigma_1^2+\log\pi_1=\frac{1}{2}\sigma_2^2+\log\pi_2 
\end{align*}
holds; thus, this model is unidentifiable.
\end{example}

%==Section 3==%
\section{Estimation}\label{Estimation}

{
To introduce model parameter estimation, we begin by defining the observed likelihood under a nonignorable missing mechanism and then discuss our estimation procedure using the FI-based approach. Additionally, we provide practical guidelines for selecting the response model in real-world applications.
}

\subsection{Observed likelihood}\label{Observed likelihood}
{
Under the model assumptions (\ref{Resp}) and (\ref{expo}), the observed likelihood is given by
\begin{equation} \label{obs_lik}
\begin{aligned}
\mathcal{L}_{\text{obs}}(\boldsymbol{\gamma},\boldsymbol{\phi}) 
&= \prod_{\delta_{i}=1} f\left(y_{i}, \delta_i=1 \mid \boldsymbol{x}_{i};\; \boldsymbol{\gamma}, \boldsymbol{\phi}\right) \prod_{\delta_{i}=0} \int f\left(y_{i}, \delta_i=0 \mid \boldsymbol{x}_{i};\; \boldsymbol{\gamma}, \boldsymbol{\phi}\right) dy_i
\\
&= \prod_{\delta_{i}=1} f\left(y_{i} \mid \boldsymbol{x}_{i}, \delta_i=1; \boldsymbol{\gamma} \right) P(\delta_i=1 \mid \boldsymbol{x}_i; \boldsymbol{\gamma}, \boldsymbol{\phi}) \prod_{\delta_{i}=0} P\left(\delta_i=0 \mid \boldsymbol{x}_{i}; \boldsymbol{\gamma}, \boldsymbol{\phi}\right)
\end{aligned}
\end{equation}
where
$\boldsymbol{\phi} := (\boldsymbol{\alpha}^\top,\beta)^\top$ and 
\[
\begin{aligned}
P\left(\delta_i=1 \mid \boldsymbol{x}_{i}; \boldsymbol{\gamma}, \boldsymbol{\phi}\right) &= \int \frac{f(y_i \mid \boldsymbol{x}_i,\delta_i=1; \boldsymbol{\gamma})}{\int f(y_i \mid \boldsymbol{x}_i,\delta_i=1; \boldsymbol{\gamma})\{P(\delta_i=1 \mid \boldsymbol{x}_i,y_i; \boldsymbol{\phi})\}^{-1}dy_i} dy_i \\
&= \frac{1}{1+ \exp\{-h(\boldsymbol{x}_i;\boldsymbol{\alpha})\} E\{\exp(-\beta y) \mid \boldsymbol{x}_i, \delta_i=1;\; \boldsymbol{\gamma}\}}.
\end{aligned}
\]
It is important to note that \(E\{\exp(-\beta y) \mid \boldsymbol{x}_i, \delta_i=1;\; \boldsymbol{\gamma}\}\) can be explicitly derived using the moment generating function of the exponential family. The details of the derivation of the observed likelihood are provided in the proof of Theorem \ref{identify multi} in the Supplementary Materials. 
}

{
A consistent estimator \(\hat{\boldsymbol{\gamma}}\) for \(\boldsymbol{\gamma}_0\), the true value of \(\boldsymbol{\gamma}\), can be obtained by solving the following score equation:
\begin{equation} \label{s1}
\mathbf{S}_1(\boldsymbol{\gamma}) = \sum_{i=1}^{n} \delta_i \mathbf{s}_{1i}(\boldsymbol{\gamma}) = \sum_{i=1}^{n} \delta_i \frac{\partial \log f(y_i \mid \boldsymbol{x}_i, \delta_i=1; \boldsymbol{\gamma})}{\partial \boldsymbol{\gamma}} = \boldsymbol{0}.
\end{equation}
Subsequently, an estimator \(\hat{\boldsymbol{\phi}}\) for \(\boldsymbol{\phi}\) can be obtained by substituting \(\hat{\boldsymbol{\gamma}}\) into (\ref{obs_lik}) and then maximizing the following function with respect to \(\boldsymbol{\phi}\):
\begin{equation} \label{obs_lik_phi}
\begin{aligned}
\mathcal{L}_{\text{obs}}(\hat{\boldsymbol{\gamma}},\boldsymbol{\phi}) \propto  \prod_{\delta_{i}=1} P(\delta_i=1 \mid \boldsymbol{x}_i; \hat{\boldsymbol{\gamma}}, \boldsymbol{\phi}) \prod_{\delta_{i}=0} P\left(\delta_i=0 \mid \boldsymbol{x}_{i}; \hat{\boldsymbol{\gamma}}, \boldsymbol{\phi}\right).
\end{aligned}    
\end{equation}
}

{
Using the estimated parameters \((\hat{\boldsymbol{\gamma}}, \hat{\boldsymbol{\phi}})\), we can compute \(\mathcal{L}_{\text{obs}}(\hat{\boldsymbol{\gamma}}, \hat{\boldsymbol{\phi}})\) and its corresponding information criterion, such as the Bayesian Information Criterion (BIC):
\begin{equation} \label{BIC}
-2 \log \mathcal{L}_{\text{obs}}(\hat{\boldsymbol{\gamma}}, \hat{\boldsymbol{\phi}}) + k \log (n),
\end{equation}
where \(k\) denotes the number of parameters, and \(n\) is the sample size. Since the respondent outcome model can be fully verified using the observed data, we can consider the respondent outcome model specification as fixed. Therefore, this information criterion will be employed to select a nonignorable nonresponse model, given a fitted respondent outcome model.
}

\subsection{Fractional imputation approach}\label{fimp}
The full data observed likelihood (\ref{obs_lik_phi}) is challenging to maximize due to the involvement of integration. \cite{riddles2016propensity} proposed a fractional imputation approach to approximate this type of integration. However, to approximate the integrations in (\ref{obs_lik_phi}), fractional imputation must be applied to both non-respondents and respondents, which would be computationally expensive. Therefore, we adopt an alternative approach where \(\hat{\boldsymbol{\phi}}\) is obtained by maximizing the response mechanism observed likelihood instead of the full data observed likelihood, as outlined in \cite{riddles2016propensity}. The goodness of fit of an estimated MNAR response model is then evaluated based on the full data observed likelihood, which remains the primary objective function to be maximized.

Specifically, \cite{riddles2016propensity} suggests obtaining \(\hat{\boldsymbol{\phi}}\) by maximizing the following response mechanism observed likelihood, which is derived directly from the response mechanism rather than from the joint probability density function of \(y\) and \(\delta\):
\begin{equation} \label{res_obs_lik}
\begin{aligned}
\mathcal{L}_{\text{res-obs}}(\boldsymbol{\gamma},\boldsymbol{\phi}) 
= \prod_{\delta_{i}=1} P(\delta_i=1 \mid \boldsymbol{x}_i, y_i; \boldsymbol{\phi}) \prod_{\delta_{i}=0} P\left(\delta_i=0 \mid \boldsymbol{x}_{i}; \boldsymbol{\gamma}, \boldsymbol{\phi}\right).
\end{aligned}
\end{equation}
This observed likelihood does not require integration for the respondents, unlike the full data observed likelihood.

However, instead of directly maximizing this observed likelihood (\ref{res_obs_lik}), we use a mean score function, where the unobserved term is approximated using the imputation approach proposed by \cite{riddles2016propensity}. To describe the proposed approach, we first define the score function \(S_i(\bm{\phi})\):
\begin{align}
    S_i(\bm{\phi}) :&= S(\bm{\alpha},\beta;\bm{x}_i,y_i,\delta_i) \label{mean score} \\
    &= \frac{\partial}{\partial (\bm{\alpha},\beta)^{\top}} \left[ \delta_i \log \pi(\bm{x}_i,y_i;\bm{\alpha},\beta) + (1-\delta_i) \log \left( 1 - \pi(\bm{x}_i,y_i;\bm{\alpha},\beta) \right) \right] \notag\\
    &= \left( \delta_i - \pi(\bm{x}_i,y_i;\bm{\alpha},\beta)  \right) \bm{z}(\bm{x},y_i;\bm{\alpha},\beta), \notag
\end{align}
where 
\begin{align*}
    \pi(\bm{x}_i,y_i;\bm{\alpha},\beta) :&= P(\delta_i=1 \mid \bm{x}_i, y_i; \bm{\alpha}, \beta), \\
   \bm{z}(\bm{x}_i,y_i;\bm{\alpha},\beta) :&= \frac{1}{\pi(\bm{x}_i, y_i;\bm{\alpha}, \beta) \left( 1-\pi(\bm{x}_i, y_i;\bm{\alpha}, \beta) \right)} \cdot \frac{\partial \pi(\bm{x}_i, y_i;\bm{\alpha}, \beta)}{\partial (\bm{\alpha}^{\top}, \beta)^{\top}}.
\end{align*}
Then, the mean score function is given by
\begin{equation}\label{msf}
\bar{S}(\bm{\gamma}, \bm{\phi}) = \sum_{i=1}^n \left[\delta_i S_i(\bm{\phi}) + (1-\delta_i) E\left\{ S_i(\bm{\phi}) \mid \bm{x}_i, \delta_i=0; \bm{\gamma}, \bm{\phi} \right\} \right].
\end{equation}
\(\hat{\bm{\gamma}}\) can be obtained by solving the score equation (\ref{s1}), and then \(\hat{\bm{\phi}}\) can be obtained by solving the mean score equation \(\bar{S}(\hat{\bm{\gamma}}, \bm{\phi}) = \bm{0}\).

However, calculating the conditional expectation in (\ref{msf}) can still be challenging due to the integration involved. To avoid this difficulty, we employ the FI-based approach proposed by \cite{kim2011parametric}. First, note that if we define the nonresponse odds ratio as \(O(\bm{x},y;\bm{\phi}) := 1/\pi(\bm{x}_i, y_i;\bm{\phi}) - 1\), the mean score function is
\begin{equation}
\label{mean score equation}
\begin{aligned}
E\left\{ S_i(\bm{\phi}) \mid \bm{x}_i, \delta_i=0; \hat{\bm{\gamma}}, \bm{\phi} \right\} &= \frac{\int S_i(\bm{\phi}) f(y \mid \bm{x}_i, \delta_i=1; \hat{\bm{\gamma}}) O(\bm{x}_i, y; \bm{\phi}) dy}{\int f(y \mid \bm{x}_i, \delta_i=1; \hat{\bm{\gamma}}) O(\bm{x}_i, y; \bm{\phi}) \, dy} \\
&= \frac{\int S_i(\bm{\phi}) f(y \mid \bm{x}_i, \delta_i=1; \hat{\bm{\gamma}}) \exp{(-\beta y)} dy}{\int f(y \mid \bm{x}_i, \delta_i=1; \hat{\bm{\gamma}}) \exp{(-\beta y)} \, dy},
\end{aligned}
\end{equation}
by the model assumption (\ref{Resp}). This expectation can be approximated by the following imputed score function:
\begin{equation} \label{isf}
\tilde{E}\left\{ S_i(\bm{\phi}) \mid \bm{x}_i, \delta_i=0; \hat{\bm{\gamma}}, \bm{\phi} \right\} := \sum_{j=1}^M w_{ij}^{*}(\bm{\phi}) S_i(\bm{\phi}; y_i^{*(j)}, \bm{x}_i),
\end{equation}
where \(w_{ij}^{*}(\bm{\phi}) := \frac{\exp{(-\beta y_i^{*(j)})}}{\sum_{k=1}^{M} \exp{(-\beta y_i^{*(k)})}}\) and \(y_i^{*(j)}\) (\(j=1, \dots, M\)) is a random sample generated from \(f(y \mid \boldsymbol{x}_i, \delta_i=1; \hat{\bm{\gamma}})\).

Finally, \(\hat{\bm{\phi}}\) can be obtained by solving the imputed score equation:
\begin{equation}\label{ise}
\begin{aligned}
S_2(\hat{\bm{\gamma}}, \bm{\phi}) :&= \sum_{i=1}^n \left[\delta_i S_i(\bm{\phi}) + (1-\delta_i) \sum_{j=1}^M w_{ij}^{*}(\bm{\phi}) S_i(\bm{\phi}; y_i^{*(j)}, \bm{x}_i) \right] \\
&= \bm{0}.
\end{aligned}
\end{equation}
This equation can be solved using an EM algorithm proposed by \cite{kim2011parametric}, with the procedure outlined as follows:

\begin{enumerate}[(Step 1)]
\item Generate \(M\) imputed values, \(y_i^{*(1)}, \dots, y_i^{*(M)}\), for each nonrespondent unit \(i\) from \(f(y \mid \boldsymbol{x}_i, \delta_i=1; \hat{\bm{\gamma}})\).
\item (E-step) With the current estimate of \(\bm{\phi}\), denoted by \(\hat{\bm{\phi}}_{(t)}\), compute the fractional weights \(w_{ij(t)}^{*} = w_{ij}^{*}(\hat{\bm{\phi}}_{(t)})\).
\item (M-step) Find \(\hat{\bm{\phi}}_{(t+1)}\) that solves the equation
   \[
   \sum_{i=1}^n \left[\delta_i S_i(\bm{\phi}) + (1-\delta_i) \sum_{j=1}^M w_{ij(t)}^{*} S_i(\bm{\phi}; y_i^{*(j)}, \bm{x}_i) \right] = \bm{0}.
   \]
\item Set $t=t+1$ and repeat (Step 2) and (Step 3) until \(\hat{\bm{\phi}}_{(t)}\) and \(\hat{\bm{\phi}}_{(t+1)}\) converge according to a specified criterion.
\end{enumerate}
If \(y\) is a discrete variable with a finite number of possible outcomes, each possible outcome of \(y\) is used as an imputed value, and the posterior probability \(P(y=y_j \mid \bm{x}_i, \delta_i=0; \hat{\bm{\gamma}}, \hat{\bm{\phi}}_{(t)})\) is used as the fractional weight \(w_{ij(t)}^*\), where \(y_j\) represents the \(j\)-th possible outcome of \(y\). After obtaining \(\hat{\bm{\phi}}\), we can estimate \(E(y)\) using the inverse probability weighted estimator:
\begin{equation} \label{ipw estimator}
\hat{\mu}_{\text{FI}} = \frac{\sum_{i=1}^n \delta_i y_i / \hat{\pi}_i}{\sum_{i=1}^n \delta_i / \hat{\pi}_i} 
\end{equation}
where \(\hat{\pi}_i = \pi(\bm{x}_i, y_i; \hat{\bm{\phi}})\).

Under certain regularity conditions, the estimator \(\hat{\bm{\phi}}\) obtained by FI is asymptotically normal, as proven by \cite{riddles2016propensity}:
\[
\sqrt{n}\left(\hat{\bm{\phi}} - \bm{\phi}_0 \right) \to N(0, \Sigma_{\bm{\phi}}),
\]
where the form of the asymptotic variance is introduced in the Supplementary Materials.

\subsection{Response Model Selection} \label{Response model selection}

Based on the information criterion and estimation procedure for the response mechanism, we propose the following practical guideline for determining the response mechanism in real-world applications:

\begin{enumerate}[(Step 1)]
    \item \label{zero step} Given fully observed cases, specify the respondents' outcome model using a GLM, \( f(y \mid \boldsymbol{x}, \delta=1; \hat{\boldsymbol{\gamma}}) \).

    \item \label{first step} Construct \(Q\) candidate response models  assuming an ignorable missing data mechanism: 
    \(\text{logit}\{P(\delta=1 \mid \boldsymbol{x})\} = h_q(\boldsymbol{x}; \hat{\boldsymbol{\alpha}}_q), \; q=1, 2, \dots, Q\). 

    By selecting among response models with an ignorable missing data mechanism, we identify candidates for the form of \(h(\boldsymbol{x}; \boldsymbol{\alpha})\) without relying on unverifiable assumptions, such as the presence of instrumental variables.
    
    \item \label{second step} Linearly add \(\beta y\) to the linear predictor \(h_q(\boldsymbol{x}; {\boldsymbol{\alpha}}_q)\), resulting in the response model candidate:
    \(\text{logit}\{P(\delta=1 \mid \boldsymbol{x}, y)\} = h_q(\boldsymbol{x}; {\boldsymbol{\alpha}}_q) + \beta y\). Estimate the response models using FI and calculate the BIC as defined in (\ref{BIC}). Select the final candidate model based on the lowest BIC.

    \item \label{third step} Test the null hypothesis that \(\beta = 0\) at a given significance level using the selected final candidate model.

    A nonidentifiable model is unlikely to be chosen, as an estimator from such a model would be inconsistent and result in poor goodness of fit. Thus, the model selected in (Step \ref{second step}) with the lowest information criterion is expected to be at least partially identifiable.
 
    \item \label{fourth step} If the null hypothesis is rejected, conclude that the response mechanism is likely nonignorable and choose the final candidate as the final response model.

    This test results offer strong evidence that the true missing data mechanism is nonignorable given our model assumptions, thereby validating the use of the selected final response model as outlined in (Step \ref{second step}).
    
    \item \label{fifth step} If the null hypothesis is not rejected, conclude that the missing mechanism can be assumed to be MAR, and use the MAR response model with the lowest BIC among the $Q$ candidates identified in (Step \ref{first step}).

    When the null hypothesis cannot be rejected, it is preferable to choose the response model with an ignorable missing data mechanism that offers the best fit. Response models with a nonignorable missing data mechanism may be unidentifiable and introduce additional variability. Therefore, if the evidence for a nonignorable mechanism is not compelling, selecting a model with an ignorable mechanism is advisable.
\end{enumerate}

As Corollary \ref{normal identification} suggests, it is preferable that the linear predictor structure of the respondents' outcome model, \(\mu(\boldsymbol{x}; \boldsymbol{\gamma})\), is more complex than \(h(\boldsymbol{x}; \boldsymbol{\alpha})\) in the response mechanism for identifiability. For this reason, we recommend using AIC in selecting the respondents' outcome model in Step \ref{zero step}, while BIC is more desirable in the response model selection stages, as AIC tends to prefer more complex models than BIC.

{Although the above procedure provides a systematic approach to selecting a final response model and deriving a single conclusion for the target parameter, it may yield biased results if the parametric models (\ref{Resp}) and (\ref{expo}) are misspecified. To mitigate potential overconfidence with misspecified models, it is advisable in real-world applications to complement our proposed procedure with sensitivity analysis, where inferences from multiple models are presented simultaneously.}

{For instance, in (Step \ref{second step}), instead of selecting a single candidate for \( h(\bm{x}; \bm{\alpha}) \), multiple candidates can be retained, and the inference results for all of them can be reported alongside the results from the MAR response model with the lowest BIC among the \( Q \) candidates identified in (Step \ref{first step}). For MAR response models, it is generally acceptable to present only the result corresponding to the model with the lowest BIC, as these models are less prone to misspecification errors compared to MNAR models.}

{If estimates exhibit low sensitivity to response model assumptions, the single conclusion derived from our proposed procedure can be adopted with greater confidence. Conversely, if high sensitivity is observed, confidence in the selected response model diminishes. In extreme cases of sensitivity, it may be prudent to refrain from selecting a single response model.}

\section{Simulation Studies}\label{Simulation}

We conduct two simulation studies to evaluate the effectiveness of the FI estimators and the proposed response model selection procedure. In the first study, we explore how the performance of our method varies under different levels of identifiability. The second study examines the robustness of our method against model misspecification, comparing it with the PPM approach introduced by \cite{andridge2011proxy}. In this context, \(\text{PPM}_{\lambda}\) denotes the PPM estimator with the sensitivity parameter \(\lambda\).
{Due to space constraints, we present only the results of the first simulation and the overview of the second simulation here.}

{
For the first simulation study, we consider two outcome models as follows: 
}
\begin{itemize}
\item[] {\textbf{Continuous case:} For \(x_1 \sim N(0.5, 0.5)\) and \(x_2 \sim N(0.5, 0.5)\), we generate a variable of interest \(y\) such that
\[
y = \theta_0 + \theta_1 x_1 + \theta_2 x_2 + \epsilon,
\]
where \(\epsilon \sim N(0, \sigma^2_\epsilon)\). 
We vary the ratio \(\theta_2/\theta_1\) from 1 to 0 to simulate different identifiability scenarios: \(\theta_2/\theta_1 = 1.0\; (\theta_1=0.57, \theta_2=0.57),\; 0.6\;(\theta_1=0.69, \theta_2=0.41),\; 0.3\;(\theta_1=0.77,\; \theta_2=0.23),\; 0.0\;(\theta_1=0.80, \theta_2=0)\). As this ratio decreases, identifiability is weakened due to the diminishing influence of the non-response instrumental variable \(x_2\). The values of \((\theta_0, \theta_1, \theta_2, \sigma_\epsilon)\) are set to ensure that the ratio between \(\theta_2\) and \(\theta_1\) matches the target ratio corresponding to each degree of identifiability. Additionally, we control \(\rho := \text{cor}(y, E(y \mid \mathbf{x}))\) at 0.8, while both \(E(y)\) and \(\text{Var}(y)\) are maintained at 0.5, regardless of the identifiability scenario.
}
\item[] {\textbf{Binary case:}  For \(x_1 \sim N(0.5, 0.25)\) and \(x_2 \sim N(0.5, 0.25)\), we generate a latent variable \(u\) such that 
\[
u = \theta_0 + \theta_1 x_1 + \theta_2 x_2 + \epsilon,
\]
where $\epsilon \sim \text{Logistic}(0,1)$ and \(y = \mathds{1}_{\{u > 0\}}\), i.e., $\text{logit}\{P(y=1 \mid \boldsymbol{x})\}=\theta_0 + \theta_1 x_1 + \theta_2 x_2$. The ratio \(\theta_2/\theta_1\) is varied in the same manner as in the continuous case. Here, \(\rho\), the correlation between \(u\) and its conditional expectation \(E(u \mid \boldsymbol{x})\), is controlled at 0.8.
The values of \((\theta_0, \theta_1, \theta_2)\) are set to ensure
\(E(y)=0.5\) and \(\text{Var}(y)=0.25\).}
\end{itemize}

{
The response mechanism is assumed to be \(\text{logit}\{P(\delta=1 \mid \boldsymbol{x}, y)\} = x_1 + y\), with a response rate of approximately 0.7 across all scenarios. For each scenario, we generate \(B = 5{,}000\) independent Monte Carlo samples, each with a sample size of \(n = 1{,}000\). 
}

{
First, we will evaluate the performance of the proposed model selection procedure. Subsequently, in each scenario, the performance of the following point estimators for \(E(y)\) will be compared based on relative bias (RB) in percentage, standard error (SE), and root mean square error (RMSE):
}
\begin{itemize}
    \item CC: Naive estimator using complete cases only.
    
    \item MAR: Inverse probability weighted estimator under a missing at random (MAR) response mechanism. The working response mechanism is selected from among the following models based on BIC:
    \begin{itemize}
        \item \(\text{logit}\{P(\delta=1 \mid \boldsymbol{x}, y)\} = \alpha_0\)
        \item \(\text{logit}\{P(\delta=1 \mid \boldsymbol{x}, y)\} = \alpha_0 + \alpha_1 x_1\)
        \item \(\text{logit}\{P(\delta=1 \mid \boldsymbol{x}, y)\} = \alpha_0 + \alpha_2 x_2\)
        \item \(\text{logit}\{P(\delta=1 \mid \boldsymbol{x}, y)\} = \alpha_0 + \alpha_1 x_1 + \alpha_2 x_2\)
    \end{itemize}
    
    \item \(\text{FI}_c\): Fractional imputation estimator described in (\ref{ipw estimator}) with the correctly specified response mechanism. Specifically, the response model employed is:
        \[
        \text{logit}\{P(\delta=1 \mid \boldsymbol{x}, y)\} = \alpha_0 + \alpha_1 x_1 + \beta y.
        \]
        The working respondents' outcome model is determined through backward elimination based on AIC. For the contiunuos case, this starts with the full model \([y \mid \boldsymbol{x}, \delta=1] \sim N(\kappa_0 + \kappa_1 x_1 + \kappa_2 x_2,\; \sigma^2)\), and for the binary case, it starts with $\text{logit}\{P(y=1 \mid \boldsymbol{x},\delta=1)\}=\kappa_0 + \kappa_1x_1 + \kappa_2x_2$. The imputation size is set to $M=100$ in the continuous case. 
        
    \item \(\text{FI}_m\): Fractional imputation estimator with a selected response mechanism selected from the following options:
        \begin{itemize}
            \item RM0: The response mechanism chosen by the MAR estimator described above.
            \item RM1: \(\text{logit}\{P(\delta=1 \mid \boldsymbol{x}, y)\} = \alpha_0 + \beta y\)
            \item RM2: \(\text{logit}\{P(\delta=1 \mid \boldsymbol{x}, y)\} = \alpha_0 + \alpha_1 x_1 + \beta y\) (true model)
            \item RM3: \(\text{logit}\{P(\delta=1 \mid \boldsymbol{x}, y)\} = \alpha_0 + \alpha_2 x_2 + \beta y\)
            \item RM4: \(\text{logit}\{P(\delta=1 \mid \boldsymbol{x}, y)\} = \alpha_0 + \alpha_1 x_1 + \alpha_2 x_2 + \beta y\)
        \end{itemize}
        Selection is based on the proposed model selection procedure with a significance level set to 10\%. The working respondents' outcome model and imputation size are set as described in \(\text{FI}_c\).
\end{itemize}

Table \ref{model_selection} presents the number of times each response model was selected as the final model. The results reveal that as $\theta_2/\theta_1$ approaches 0, the true response model (RM2) is selected less often, while RM1 is chosen more frequently, especially in the continuous case when \(\theta_2/\theta_1 = 0\). This outcome demonstrates that our model selection procedure effectively excludes unidentifiable models by comparing the goodness-of-fit among candidates. Conversely, when identifiability is strong, our method successfully identifies the true response mechanism. Additional discussion on the response model selection is provided in the Supplementary Materials.

\begin{table}
\caption{\label{model_selection} The number of times each response model was selected}
\centering
\begin{tabular}{@{}cclrrrrr@{}}
\toprule
$y$ & $\theta_2/\theta_1$ &  & RM0 & RM1 & RM2 & RM3 & RM4 \\ \midrule
\multirow{4}{*}{Continuous} & 1.0 &  & 1 & 0 & 4898 & 101 & 0 \\
 & 0.6 &  & 87 & 115 & 4209 & 589 & 0 \\
 & 0.3 &  & 1059 & 2388 & 703 & 850 & 0 \\
 & 0.0 &  & 0 & 4934 & 0 & 66 & 0 \\ \midrule
\multirow{4}{*}{Binary} & 1.0 &  & 222 & 21 & 4642 & 93 & 22 \\
 & 0.6 &  & 854 & 265 & 3708 & 160 & 13 \\
 & 0.3 &  & 2051 & 1154 & 1671 & 117 & 7 \\
 & 0.0 &  & 2364 & 2064 & 544 & 23 & 5 \\ \bottomrule
\end{tabular}
\end{table}

Table \ref{sim1_point} presents the Monte Carlo means of RB, SE and RMSE for the four estimators on \(E(y)\) and $\beta$.
In estimating \(E(y)\), the CC and MAR exhibit significant bias across all scenarios. In contrast, \(\text{FI}_c\) and \(\text{FI}_m\) are almost unbiased in scenarios where \(\theta_2/\theta_1 >0\). For the scenario with \(\theta_2/\theta_1 = 0\), both fractional imputation estimators provide slight bias in the case of a continuous outcome variable; however,  the biases are comparable to that of the MAR estimator. In the binary outcome case, both estimators remain unbiased, because the logistic outcome model may still be identifiable as discussed in Example \ref{binomial Example}. For a response model parameter $\beta$, we report results only for \(\text{FI}_c\), as the selected response model in \(\text{FI}_m\) may not represent the true response mechanism. When the response model is identifiable, \(\text{FI}_c\) consistently provides nearly unbiased estimates.

Also note that \(\text{FI}_m\) tends to produce more bias and generally exhibits greater variability as identifiability weakens. When identifiability is strong, \(\text{FI}_m\) can accurately select the correct response model, leading to reliable estimates. However, as identifiability diminishes, the model selection process becomes less effective, increasing the chance of choosing an incorrect model. This misspecification introduces errors in the parameter estimation, resulting in greater bias and variability in \(\text{FI}_m\) compared to \(\text{FI}_c\), which relies on a correctly specified model from the outset.

\begin{table}[!htp]
\caption{\label{sim1_point}Monte Carlo means of Relative Bias (RB, \%), Standard Error (SE), and Root Mean Square Error (RMSE) of the point estimators in Simulation I}
\centering
\resizebox{\textwidth}{!}{%
\begin{tabular}{@{}ccllrrrlrrrlrrrlrrr@{}}
\toprule
\multicolumn{1}{l}{} &  & \multicolumn{1}{c}{} & \multicolumn{1}{c}{} & \multicolumn{3}{c}{$\theta_2/\theta_1=1.0$} & \multicolumn{1}{c}{} & \multicolumn{3}{c}{$\theta_2/\theta_1=0.6$} & \multicolumn{1}{c}{} & \multicolumn{3}{c}{$\theta_2/\theta_1=0.3$} & \multicolumn{1}{c}{} & \multicolumn{3}{c}{$\theta_2/\theta_1=0.0$} \\ \cmidrule(lr){5-7} \cmidrule(lr){9-11} \cmidrule(lr){13-15} \cmidrule(l){17-19} 
$y$ & Parameter & Method &  & RB & SE & RMSE &  & RB & SE & RMSE &  & RB & SE & RMSE &  & RB & SE & RMSE \\ \midrule
\multirow{5}{*}{Continuous} & \multirow{4}{*}{$E(y)$} & CC &  & 38.6 & 0.025 & 0.195 &  & 41.3 & 0.025 & 0.208 &  & 43.1 & 0.025 & 0.217 &  & 43.8 & 0.025 & 0.221 \\
 &  & MAR &  & 11.0 & 0.027 & 0.061 &  & 11.5 & 0.028 & 0.064 &  & 12.0 & 0.028 & 0.066 &  & 11.2 & 0.028 & 0.062 \\
 &  & $\text{FI}_c$ &  & 0.2 & 0.032 & 0.032 &  & 0.1 & 0.035 & 0.035 &  & -0.5 & 0.044 & 0.044 &  & 9.5 & 0.089 & 0.101 \\
 &  & $\text{FI}_m$ &  & -0.1 & 0.033 & 0.033 &  & -1.5 & 0.045 & 0.046 &  & -6.1 & 0.068 & 0.075 &  & -13.6 & 0.051 & 0.085 \\ \cmidrule(l){2-19} 
 & $\beta$ & $\text{FI}_c$ &  & 1.3 & 0.211 & 0.212 &  & 2.4 & 0.291 & 0.292 &  & 8.1 & 0.550 & 0.556 &  & -83.2 & 1.574 & 1.780 \\ \midrule
\multirow{5}{*}{Binary} & \multirow{4}{*}{$E(y)$} & CC &  & 19.7 & 0.018 & 0.100 & \multicolumn{1}{r}{} & 21.1 & 0.018 & 0.107 & \multicolumn{1}{r}{} & 22.0 & 0.018 & 0.112 & \multicolumn{1}{r}{} & 22.4 & 0.018 & 0.114 \\
 &  & MAR &  & 8.4 & 0.019 & 0.046 & \multicolumn{1}{r}{} & 8.4 & 0.018 & 0.046 & \multicolumn{1}{r}{} & 8.0 & 0.018 & 0.044 & \multicolumn{1}{r}{} & 7.5 & 0.018 & 0.042 \\
 &  & $\text{FI}_c$ &  & 0.1 & 0.020 & 0.020 & \multicolumn{1}{r}{} & 0.2 & 0.022 & 0.022 & \multicolumn{1}{r}{} & 0.4 & 0.025 & 0.025 & \multicolumn{1}{r}{} & 1.6 & 0.035 & 0.036 \\
 &  & $\text{FI}_m$ &  & 0.1 & 0.024 & 0.024 & \multicolumn{1}{r}{} & 0.4 & 0.029 & 0.029 & \multicolumn{1}{r}{} & 0.7 & 0.037 & 0.038 & \multicolumn{1}{r}{} & -0.1 & 0.042 & 0.042 \\ \cmidrule(l){2-19} 
 & $\beta$ & $\text{FI}_c$ &  & 0.9 & 0.314 & 0.314 & \multicolumn{1}{r}{} & 0.7 & 0.400 & 0.400 & \multicolumn{1}{r}{} & -1.3 & 0.569 & 0.569 & \multicolumn{1}{r}{} & -15.5 & 0.844 & 0.858 \\ \bottomrule
\end{tabular}%
}
\end{table}

Table \ref{var_sim1} summarizes the Monte Carlo means of RB of variance estimators and CR of 95\% confidence intervals under correct specification of the response model. The linearised variance estimator in the Supplementary Materials is employed for the \(\text{FI}_c\) method. The \(\text{FI}_m\) method is excluded from this variance estimation due to variability in the selected response model across Monte Carlo samples.  As anticipated, the CR performance of both the CC and MAR estimators is notably poor, attributable to the nonignorable missing data mechanism. In contrast, The \(\text{FI}_c\) estimator demonstrates strong performance in terms of both RB and CR in scenarios where the model is identifiable. However, substantial biases arise when the response model is not identifiable, particularly in the case of a continuous outcome variable.

\begin{table}[!htp]
\caption{\label{var_sim1}
Monte Carlo means of Relative Bias (RB, \%) of variance estimators and Coverage Rate (CR, \%) for 95\% confidence intervals}
\centering
\resizebox{\textwidth}{!}{%
\begin{tabular}{@{}cclcrrrrrrrrrrr@{}}
\toprule
\multicolumn{1}{l}{} &  & \multicolumn{1}{c}{} &  & \multicolumn{2}{c}{$\theta_2/\theta_1=1.0$} & \multicolumn{1}{c}{} & \multicolumn{2}{c}{$\theta_2/\theta_1=0.6$} & \multicolumn{1}{c}{} & \multicolumn{2}{c}{$\theta_2/\theta_1=0.3$} & \multicolumn{1}{c}{} & \multicolumn{2}{c}{$\theta_2/\theta_1=0.0$} \\ \cmidrule(lr){5-6} \cmidrule(lr){8-9} \cmidrule(lr){11-12} \cmidrule(l){14-15} 
Outcome & Parameter & Method &  & RB & CR &  & RB & CR &  & RB & CR &  & RB & CR \\ \midrule
\multirow{4}{*}{Continuous} & \multirow{3}{*}{Var$(\hat{\mu})$} & CC &  & 0.4 & 0.0 &  & -0.6 & 0.0 &  & -1.3 & 0.0 &  & -2.0 & 0.0 \\
 &  & MAR &  & -0.8 & 42.5 &  & -2.6 & 41.6 &  & 0.8 & 39.8 &  & 1.2 & 45.1 \\
 &  & $\text{FI}_{c}$ &  & 1.4 & 95.6 &  & -0.1 & 94.6 &  & -2.9 & 94.9 &  & 162.8 & 96.8 \\ \cmidrule(l){2-15} 
 & Var$(\hat{\beta})$ & $\text{FI}_{c}$ &  & -1.6 & 95.0 &  & -2.2 & 95.0 &  & -5.6 & 95.1 &  & 157.5 & 97.7 \\ \midrule
\multirow{4}{*}{Binary} & \multirow{3}{*}{Var$(\hat{\mu})$} & CC & \multicolumn{1}{l}{} & 0.3 & 0.1 &  & 1.6 & 0.0 &  & 2.2 & 0.0 &  & 0.7 & 0.0 \\
 &  & MAR & \multicolumn{1}{l}{} & -5.9 & 37.5 &  & -0.9 & 36.1 &  & 1.3 & 39.2 &  & -1.3 & 44.1 \\
 &  & $\text{FI}_{c}$ & \multicolumn{1}{l}{} & 1.8 & 95.4 &  & 4.5 & 95.3 &  & 2.3 & 95.1 &  & -10.0 & 94.0 \\ \cmidrule(l){2-15} 
 & Var$(\hat{\beta})$ & $\text{FI}_{c}$ & \multicolumn{1}{l}{} & 1.5 & 95.5 &  & 2.7 & 95.9 &  & 0.3 & 95.3 &  & -12.0 & 94.4 \\ \bottomrule
\end{tabular}%
}
\end{table}

%{Finally, we provide a brief overview of the second simulation study. For details, please refer to the Supplementary Material.The second simulation study evaluates the robustness of our proposed FI estimators against model misspecification, comparing them with PPM and MAR approaches. Table A1 in Supplementary Material demonstrates that FI estimators produce nearly unbiased estimates across various scenarios, whereas PPM estimators exhibit significant bias when their assumptions are violated, particularly for non-normal outcome models. The MAR estimators also show bias under nonignorable missing mechanisms.Table A2 in Supplementary Material highlights that FI estimators remain robust even under response model misspecification, with smaller biases compared to PPM estimators. However, exceptions occur in specific cases (e.g., Probit or Log-log link functions), where the bias of \(\text{FI}_m\) estimators reaches up to 27.7\%.}

{Finally, we provide a brief overview of the second simulation study.
In the second simulation study, we evaluate the robustness of the proposed FI estimators under various scenarios of model misspecification, comparing them with the PPM and MAR estimators.
Specifically, we evaluate the performance of these methods in estimating $E(y)$ across different outcome and response model conditions.
The results from outcome model misspecification demonstrate that FI estimators produce nearly unbiased estimates of $E(y)$ in all scenarios, regardless of the true outcome model.
In contrast, PPM estimators show significant biases when their underlying assumptions do not align with the true data-generating process.
For example, biases in the PPM estimators are particularly pronounced under non-normal outcome models.
In the case of response model misspecification, the FI estimators remain robust in most scenarios.
However, some exceptions are observed, particularly for Probit and Log-log link functions, where the \(\text{FI}_m\) estimator exhibited biases of up to 27.7\%.
Despite these cases, FI estimators generally outperform PPM and MAR estimators, highlighting their robustness under a wide range of conditions.
The strength of the proxy (\(\rho\)) also play a critical role in estimator performance.
When the proxy is weak (\(\rho = 0.5\)), biases and RMSE values increase across all methods.
Conversely, strong proxy (\(\rho = 0.8\)) significantly reduce these metrics, emphasizing the importance of incorporating high-quality proxies in real-world studies.
Further details on the simulation settings, as well as additional results, can be found in the Supplementary Material.
}

\section{Real Data Analysis} \label{Real Data}

{This section presents the analysis results of real-world data employing the proposed methodology. The first case study utilizes survey data to estimate the mean of a continuous outcome variable. The second case involves predicting election outcomes using exit poll data, with a focus on categorical outcome variables. Detailed discussions are provided in the subsequent subsections.
}

\subsection{Korean Survey of Household Finances and Living Conditions Data}\label{Real Data 1}

We analyse the Korean Survey of Household Finance and Living Conditions data collected by Korea Statistics (\url{https://mdis.kostat.go.kr/index.do}) in 2021 via researcher interviews and Internet-based surveys. This dataset encompasses 160 variables, including household income, expenditure, and the age of the household head. From a total of 18,187 households, we selected 17,632 units, specifically those with incomes surpassing the legal minimum living cost and expenditures exceeding zero.

{
The primary variable of interest is the logarithm-transformed annual \textit{household income}, denoted by \(y\). Logarithm-transformed \textit{consumption expenditure} serves as the first auxiliary variable, denoted by \(x_1\), while logarithm-transformed \textit{non-consumption expenditure} is represented by \(x_2\). The \textit{age} of the household head, scaled by division by ten, is the third covariate, represented by \(x_3\). All of these variables are considered continuous. The response indicator \(\delta\) is defined such that it assumes a value of 1 for male household heads and 0 otherwise. In this hypothetical scenario, where \(y\) is unobserved for households with a female head, the respondent pool consists of 13,048 out of 17,632 units, resulting in a response rate of 74\%.
}

{
To obtain \(\text{FI}\) estimates, we first fit the respondents' outcome model using backward elimination, guided by AIC, beginning with a saturated model that includes following predictors: \(\boldsymbol{x} = (x_1, x_2, x_3, x_1^2, x_2^2, x_3^2, x_1x_2, x_1x_3, x_2x_3, x_1x_2x_3)\). 
The resulting model is:
\[
\begin{aligned}
\relax [y \mid \boldsymbol{x}, \delta=1; \hat{\boldsymbol{\gamma}}] \sim N&(2.47 + 0.73x_1 + 0.11x_2 - 0.03x_1^2 + 0.04x_2^2 \\
&\quad - 0.01x_3^2 - 0.02x_1x_2 + 0.04x_1x_3 - 0.02x_2x_3,\; 0.37^2),
\end{aligned}
\]
with an R-squared value of 0.75. This fitted respondents' outcome model is used to generate $M=200$ imputed values for each non-respondent. Additionally, the fitted values from this regression model serve as the proxy variable for the $\text{PPM}_{\lambda}$ methods. This proxy variable exhibits a high correlation of 0.87 with 
an outcome variable $y$.
}

{
For the response model, we use a simple logistic regression that incorporates first-order terms for the $\text{FI}_{c}$ estimator,
while the $\text{FI}_{m}$ uses a response model selected based on the practical guidelines outlined in Section \ref{Response model selection}.  The estimated response model for $\text{FI}_{c}$ is:
\[
\text{logit}\{P(\delta=1\mid \boldsymbol{x}, y; \hat{\boldsymbol{\phi}}_c)\}=-13.34+126x_1+0.06x_2+0.15x_3+0.46y.
\]
}

{
For response model selection of the \(\text{FI}_m\) estimator, we first use a best subset model selection approach to identify candidate response models with the lowest BIC values. In this study, we select \(Q=5\) response model candidates, and the response model with the lowest BIC value is fitted as follows:
\begin{equation}
\label{mar}
\begin{aligned}
\text{logit}\{P(\delta=1 \mid \boldsymbol{x}; \hat{\boldsymbol{\alpha}})\} &= 18.85 - 7.47x_2 - 6.74x_3 - 0.36x_1^2 + 0.03x_3^2 \\
&\quad + 1.05x_1x_2 + 0.91x_1x_3 + 0.98x_2x_3 - 0.14x_1x_2x_3.
\end{aligned}    
\end{equation}
This response model is used to compute the MAR estimate.
}

{
After identifying five response models with an ignorable missing data mechanism through the model selection procedure described above, we generate candidate response models for the \(\text{FI}_{m}\) estimator by adding an additional predictor term \(\beta \times y\), resulting in the model form \(\text{logit}\{P(\delta=1 \mid \boldsymbol{x}, y)\} = h_q(\boldsymbol{x}; \boldsymbol{\alpha}_q) + \beta y\), where \(q = 1, \dots, 5\). Among these five models with a nonignorable missing data mechanism, the one with the lowest BIC value, calculated using formula (\ref{BIC}), is selected as a final candidate. This candidate is chosen as the final response model if the \(p\)-value for the hypothesis test of \(\beta = 0\) is less than the significance level of 0.1. If not, we revert to the response model (\ref{mar}), which assumes an ignorable missing data mechanism.
}

\begin{table}[h]
\centering
\caption{BIC and p-values of $\beta$ for five response model candidates}
\label{real_conti_model}
\resizebox{\textwidth}{!}{%
\begin{tabular}{@{}cccccccccccccc@{}}
\toprule
                                            & $x_1$ & $x_2$ & $x_3$ & $x_1^2$ & $x_2^2$ & $x_3^2$ & $x_1x_2$ & $x_1x_3$ & $x_2x_3$ & $x_1x_2x_3$ & BIC   & p-value of $\beta$  \\ \midrule
$h_1(\boldsymbol{x};\boldsymbol{\alpha}_1)$ & x     & o     & o     & o       & x       & o       & o        & o        & o        & o           & 16756 & 0.08                         \\
$h_2(\boldsymbol{x};\boldsymbol{\alpha}_2)$ & o     & o     & o     & o       & x       & o       & o        & o        & o        & o           & 16762 & 0.09                         \\
$h_3(\boldsymbol{x};\boldsymbol{\alpha}_3)$ & x     & o     & o     & o       & x       & x       & o        & o        & o        & o           & 16764 & 0.20                         \\
$h_4(\boldsymbol{x};\boldsymbol{\alpha}_4)$ & x     & o     & o     & o       & o       & o       & o        & o        & o        & o           & 16765 & 0.88                         \\
$h_5(\boldsymbol{x};\boldsymbol{\alpha}_5)$ & x     & o     & o     & o       & o       & x       & o        & o        & o        & o           & 16769 & 0.94                         \\ \midrule
o: selected; x: not selected                &       &       &       &         &         &         &          &          &          &             &       &                    &               
\end{tabular}%
}
\end{table}

\begin{figure}[h]
    \centering
    \caption{Point estimates and 95\% confidence intervals for $E(y)$ by estimator}
    \includegraphics[width=\linewidth]{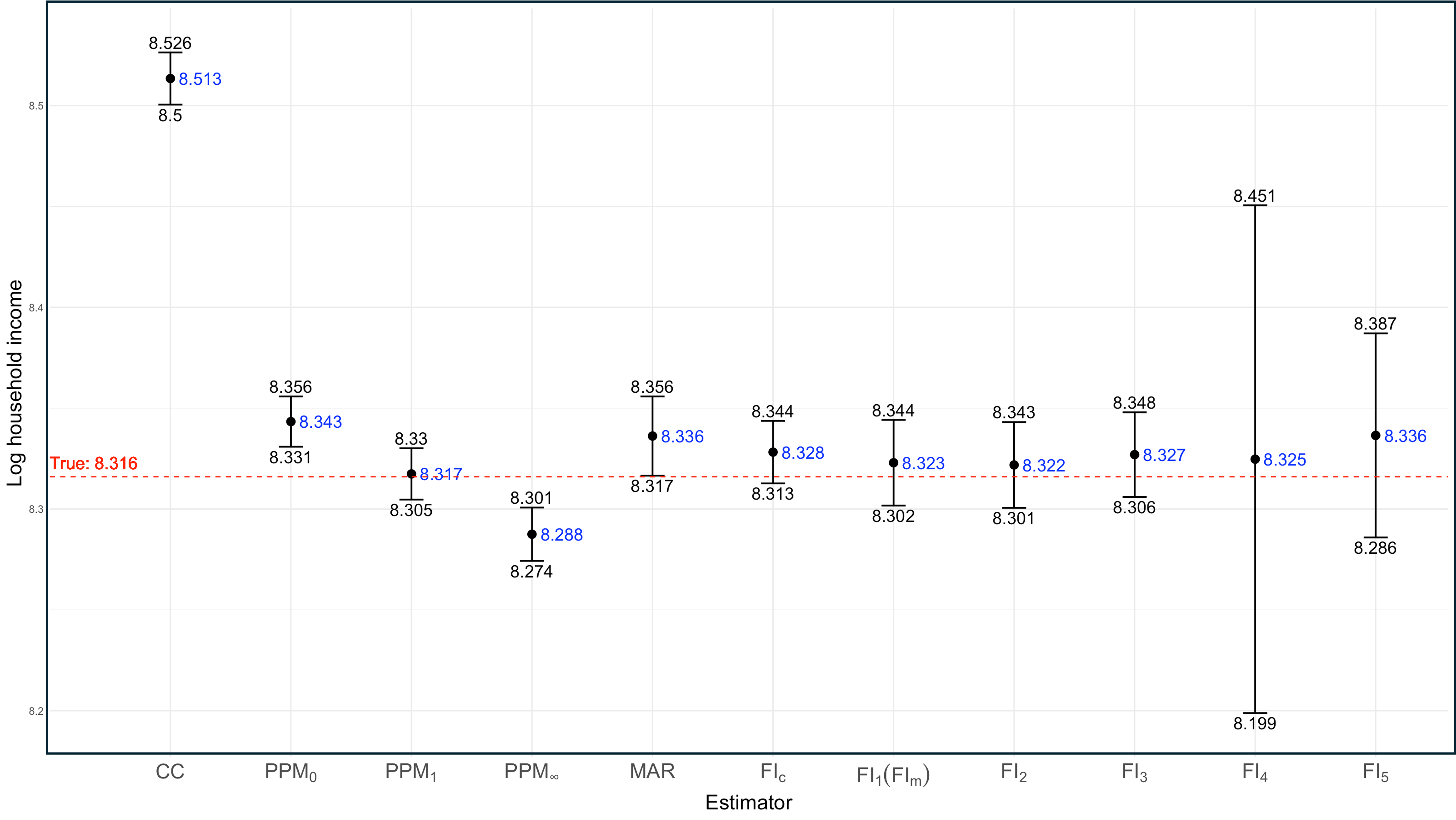}
    \label{real_conti_sum}
\end{figure}

% \begin{table}[h]
% \centering
% \caption{Point estimates and 95\% confidence intervals for $E(y)$}
% \label{real_conti_sum}
% \begin{tabular}{@{}lccc@{}}
% \toprule
% Method & Point Estimate & 95\% C.I. & Length of C.I. \\ \midrule
% True & 8.316 & - & - \\
% CC & 8.513 & (8.500, 8.526) & 0.026 \\
% $\text{PPM}_0$ & 8.343 & (8.331, 8.356) & 0.025 \\
% $\text{PPM}_1$ & 8.317 & (8.305, 8.330) & 0.025 \\
% $\text{PPM}_{\infty}$ & 8.288 & (8.274, 8.301) & 0.026 \\
% MAR & 8.336 & (8.317, 8.356) & 0.039 \\
% $\text{FI}_c$ & 8.328 & (8.313, 8.344) & 0.031 \\
% $\text{FI}_1(\text{FI}_m)$ & 8.323 & (8.302, 8.344) & 0.042 \\
% $\text{FI}_2$ & 8.322 & (8.301, 8.343) & 0.042 \\
% $\text{FI}_3$ & 8.327 & (8.306, 8.348) & 0.042 \\
% $\text{FI}_4$ & 8.325 & (8.199, 8.451) & 0.252 \\
% $\text{FI}_5$ & 8.336 & (8.286, 8.387) & 0.101 \\ \bottomrule
% \end{tabular}
% \end{table}

The five response model candidates, along with their BIC values and p-values for the hypothesis test are summarized in Table \ref{real_conti_model}.
Given that the p-value for the hypothesis test of the best candidate is 0.08, we adopt \(\text{logit}\{P(\delta=1 \mid \boldsymbol{x}, y)\} = h_1(\boldsymbol{x}; {\boldsymbol{\alpha}}_1) + \beta y\) as our final response model. The model is estimated as follows:
\[
\begin{aligned}
\text{logit}\{P(\delta=1 \mid \boldsymbol{x}, y; \hat{\boldsymbol{\phi}}_m)\} &= 16.28 - 7.34x_2 - 6.59x_3 - 0.35x_1^2 + 0.04x_3^2 \\
&\quad + 1.01x_1x_2 + 0.87x_1x_3 + 0.98x_2x_3 - 0.13x_1x_2x_3 + 0.36y.
\end{aligned}
\]
Note that, in general cases, the predictor terms in \(h(\bm{x};\bm{\alpha})\) of the \(\text{FI}_m\) estimator may differ from those of the MAR estimator due to changes in BIC values after the additional predictor term \(\beta \times y\) is included. However, in this example, both the MAR and \(\text{FI}_m\) estimators share the same predictor terms as in \(h_1(\bm{x}; \bm{\alpha}_1)\), as it remains the best model even after adding \(\beta \times y\).

{Figure \ref{real_conti_sum} displays the point estimates and 95\% confidence intervals for $E(y)$ by estimator. Following the proposal by \cite{andridge2011proxy}, we consider $\lambda \in \{0,1,\infty\}$ for the $\text{PPM}_{\lambda}$ methods. {The $\text{FI}_q$ estimator for $q=1,...,5$ employs the working response model $\text{logit}\{P(\delta=1 \mid \bm{x}, y)\} = h_q(\bm{x}; \bm{\alpha}_q) + \beta y$. Notably, since $\text{FI}_1$ utilizes the response model selected by our proposed procedure, it is denoted as $\text{FI}_m$ in parentheses.} As illustrated, the naive CC estimate significantly deviates from the true value. While the MAR estimator reduces this bias somewhat, its confidence interval still does not encompass the true value. In contrast, the \(\text{FI}_m\) estimate is closer to the true value, with the true value located near the center of its confidence interval. \(\text{PPM}_1\) also accurately estimates the true value, and its confidence interval is narrower than that of \(\text{FI}_m\), indicating better efficiency. However, unlike \(\text{FI}_m\), which decisively selects the response model, the PPM framework cannot extract any information from the data to select the value of the sensitivity parameter \(\lambda\). Considering this uncertainty in the choice of \(\lambda\), the confidence interval for the PPM method may ideally combine the intervals from all three \(\text{PPM}_{\lambda}\) estimators, resulting in a wider interval than that of \(\text{FI}_m\).  Additionally, in this application, the \(\text{FI}_c\) estimator provides a noteworthy estimate even when using a naively specified logistic response model.
}

{Additionally, the confidence intervals from the $\text{FI}_4$ and $\text{FI}_5$ estimators are unreasonably large compared to those from the other estimators, indicating that these two estimators rely on unidentifiable models. By conducting sensitivity analysis using only estimates from identifiable models (MAR, $\text{FI}_c$, $\text{FI}_1$, $\text{FI}_2$, $\text{FI}_3$), we observe that the location and length of the confidence intervals remain consistent across various response model assumptions. This consistency arises from the high goodness of fit of the outcome model for respondents, which allows the proxy variable to provide strong information about the missing $y$. Consequently, in this real-world application example, we can confidently adopt the single inference derived from the selected response model.}

\subsection{Exit Poll Data}
\label{Real Data 2}
We also applied our methodological framework to exit poll data from the 19th South Korean legislative election held in 2012. This dataset was previously analyzed by \cite{riddles2016propensity}, who demonstrated the superior predictive accuracy achieved by accounting for a nonignorable response mechanism within this dataset. 
See \cite{riddles2016propensity} for more details on data collection and background information.

{
In this election, although multiple candidates contested each district, the competition was primarily between two major candidates vying for a single seat per district. Therefore, the analysis was simplified to consider only the votes for these two main contenders, with voter preference denoted by \(y\) as a binary variable: 1 indicating support for the first major candidate and 0 for the second. Accordingly, District 39, Gwanak-eul, which had three major candidates competing, was excluded from the analysis, leaving 47 out of 48 districts in Seoul for analysis. Note that the two major parties competing for a single seat could vary across districts.
}

The first covariate, \textit{gender}, is a binary variable (\(z_1\)), with 1 representing males and 0 representing females. The second auxiliary variable, \textit{age}, is an ordered categorical variable grouped into five categories (20s, 30s, 40s, 50s, and 60+), encoded by \(z_2\), a natural number ranging from 1 to 5. Table \ref{data structure} illustrates the data structure for District 47, Gangdong-Gap, as an example.

\begin{table}[h]
\caption{\label{data structure}Data structure of the exit poll data for District 47}
\centering
\begin{tabular}{@{}ccrrrr@{}}
\toprule
Gender & Age group & Voted A & Voted B & Refusal & Total \\ \midrule
Male & 20s & 93 & 115 & 28 & 236 \\
 & 30s & 104 & 233 & 82 & 419 \\
 & 40s & 146 & 295 & 49 & 490 \\
 & 50s & 265 & 228 & 86 & 579 \\
 & 60+ & 295 & 122 & 88 & 505 \\ \midrule
Female & 20s & 106 & 159 & 62 & 327 \\
 & 30s & 129 & 242 & 70 & 441 \\
 & 40s & 170 & 262 & 69 & 501 \\
 & 50s & 284 & 137 & 86 & 507 \\
 & 60+ & 217 & 81 & 125 & 423 \\ \midrule
Total &  & 1809 & 1874 & 745 & 4428 \\ \bottomrule
\end{tabular}
\end{table}

{
To efficiently capture how \(y\) varies with changes in ordered categories using a linear combination of orthogonal polynomials, 
we employ polynomial coding on the ordinal information of \textit{age}.
This process is implemented with the R software package \citep{fox2018r}. Given that our dataset categorizes age into five groups, we include polynomial terms up to the fourth degree. Polynomial coding assigns specific polynomial values to each unit based on its age group membership, as shown in Table \ref{poly_contrast}. These values are then used as predictors in the model. In this context, \(x_{ba}\) denotes the value of the \(b\)-th (\(b=1,2,3,4\)) order polynomial function for the \(a\)-th (\(a=1,2,3,4,5\)) age group.
}

\begin{table}
\caption{\label{poly_contrast}
Polynomial contrasts up to the 4th degree}
\centering
\begin{tabular}{@{}lrrrr@{}}
\toprule
 & Linear & Quadratic & Cubic & Quartic \\ \midrule
20s & -0.63 & 0.53 & -0.32 & 0.12 \\
30s & -0.32 & -0.27 & 0.63 & -0.48 \\
40s & 0.00 & -0.53 & 0.00 & 0.72 \\
50s & 0.32 & -0.27 & -0.63 & -0.48 \\
60+ & 0.63 & 0.53 & 0.32 & 0.12 \\ \bottomrule
\end{tabular}
\end{table}

{
To obtain the prediction estimates, both the respondents' outcome model and the response models are repeatedly specified within each district. For the FI estimators, the respondents' outcome model is determined by applying backward elimination to the following saturated model: 
\[
\label{exitpoll_outcome}
\begin{aligned}
\text{logit}\{P(y=1 \mid z_1,z_2,\delta=1; \boldsymbol{{\gamma}})\} &= \sum_{a=1}^{5}I(z_2=a) (\kappa_0 + \kappa_1x_{1a} + \kappa_2x_{2a} + \kappa_3x_{3a} + \kappa_4x_{4a} \\
& \quad + \kappa_5z_1 + \kappa_6z_1x_{1a} + \kappa_7z_1x_{2a} + \kappa_8z_1x_{3a} + \kappa_9z_1x_{4a}).
\end{aligned}
\]
}

{
For the response model of the \(\text{FI}_c\) estimator, we simply assume the following response model across all 47 districts:
\[
\text{logit}\{P(\delta=1 \mid z_1, z_2, y; \boldsymbol{\phi})\} = \sum_{a=1}^{5} I(z_2=a) (\alpha_0 + \alpha_1 x_{1a} + \alpha_2 z_1 + \alpha_3 z_1 x_{1a}) + \beta y.
\]
For the $\text{FI}_{m}$ estimator, we use the same model selection procedure discussed at the previous application, starting with the following saturated response model: 
\[
\label{exitpoll_response}
\begin{aligned}
\text{logit}\{P(\delta=1 \mid z_1, z_2; \boldsymbol{{\alpha}})\} &= \sum_{a=1}^{5}I(z_2=a) (\alpha_0 + \alpha_1x_{1a} + \alpha_2x_{2a} + \alpha_3x_{3a} + \alpha_4x_{4a} \\
& \quad + \alpha_5z_1 + \alpha_6z_1x_{1a} + \alpha_7z_1x_{2a} + \alpha_8z_1x_{3a} + \alpha_9z_1x_{4a}).
\end{aligned}
\]
According to the response model selection results, ignorable missing data mechanism models are applied in 37 districts, whereas nonignorable missing data mechanism models are used in 10 districts.
}

% \begin{table}
% \caption{\label{CI_ExitPoll_tab}Point estimates and confidence intervals for \(E(y)\)}
% \centering
% \begin{tabular}{@{}cclcc@{}}
% \toprule
% District & True & Method & Point Estimate & 95\% C.I.  \\ \midrule
% \multirow{4}{*}{22} & \multirow{4}{*}{0.506} & CC & 0.479 & (0.464, 0.494)  \\
%  &  & MAR & 0.487 & (0.472, 0.502)  \\
%  &  & $\text{FI}_c$ & 0.542 & (0.500, 0.585) \\
%  &  & $\text{FI}_m$ & 0.543 & (0.500, 0.586) \\ \midrule
% \multirow{4}{*}{47} & \multirow{4}{*}{0.519} & CC & 0.491 & (0.475, 0.507)  \\
%  &  & MAR & 0.494 & (0.478, 0.511)  \\
%  &  & $\text{FI}_c$ & 0.534 & (0.485, 0.584)  \\
%  &  & $\text{FI}_m$ & 0.530 & (0.498, 0.561) \\ \bottomrule
% \end{tabular}
% \end{table}

\begin{figure}[h]
    \centering
    \caption{Point estimates and 95\% confidence intervals for \(E(y)\) in Districts 22 and 47}
    \includegraphics[width=\linewidth]{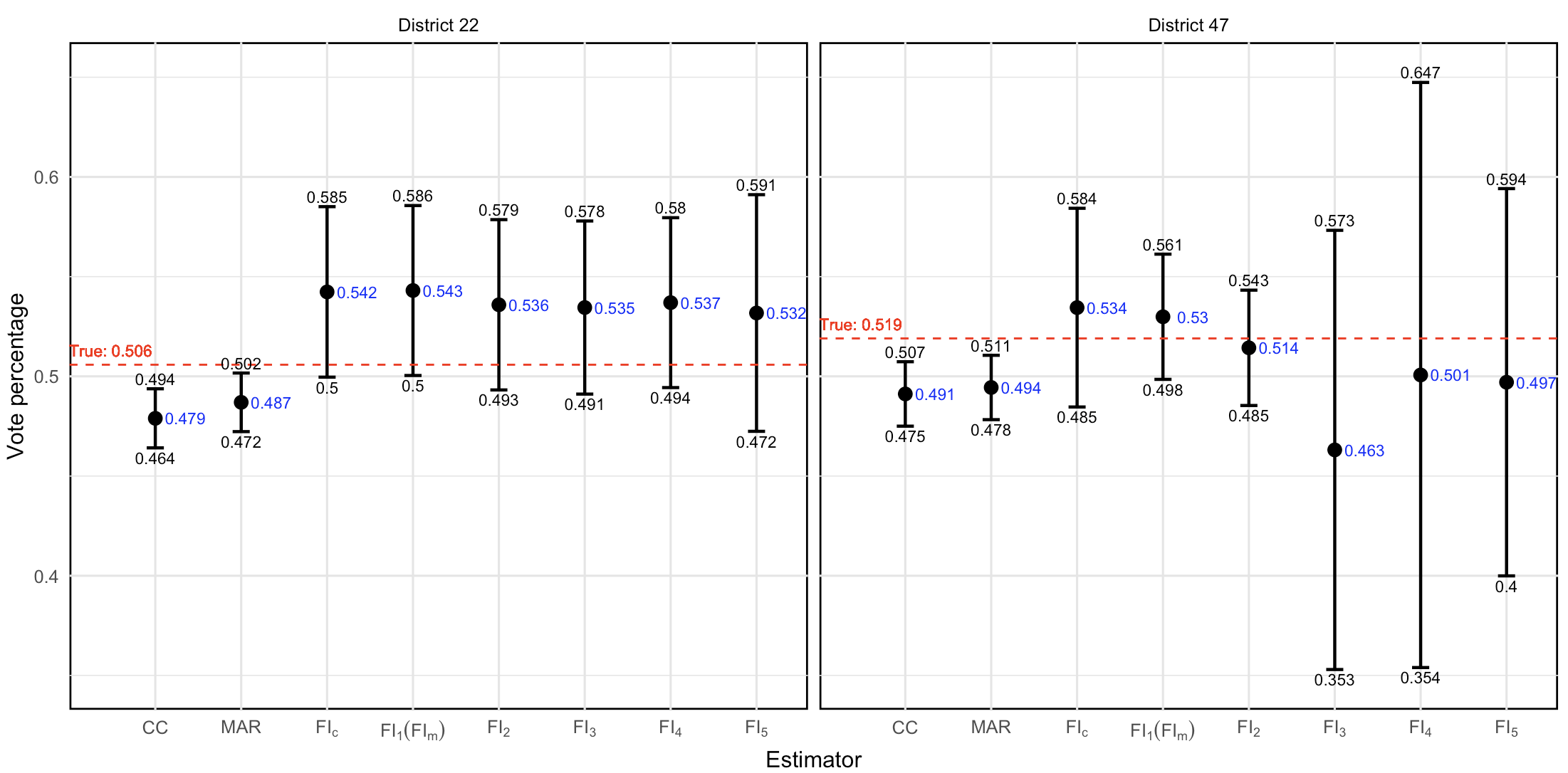}
    \label{CI_ExitPoll_tab}
\end{figure}

{
Figure \ref{CI_ExitPoll_tab} presents the point estimates and 95\% confidence intervals for \(E(y)\) in two example districts, where nonignorable response models were used to derive the $\text{FI}_{m}$ estimates. Since all variables in this dataset are categorical, the PPM method is excluded from this analysis. Sensitivity analysis reveals that $\text{FI}_3$, $\text{FI}_4$, and $\text{FI}_5$ rely on unidentifiable models in District 47, justifying their exclusion from the analysis.}

{
When comparing the CC, MAR, $\text{FI}_c$, and $\text{FI}_m$ estimators, only the 95\% confidence intervals from the FI estimators include the true values, whereas those from the CC and MAR estimators do not. While the proposed $\text{FI}_m$ estimator partially adjusts for nonignorable nonresponse bias, caution is required when applying the model selection procedure to this dataset in real-world situation where the true value is unknown. The results demonstrate high sensitivity to response model assumptions, particularly in comparisons between MAR estimators and FI estimators assuming MNAR missing mechanisms.}

\begin{table}[h]
\caption{\label{election prediction}Predicted parliamentary seats, average bias (Avg.Bias), and the coverage results with 95\% and 99\% confidence intervals. `Length' denotes the average length of confidence intervals and `No. of dist.' denotes the number of districts where the confidence interval contains true value. Numbers in parentheses indicate results for the 10 districts where the MNAR-type response model was selected.}
\centering
\resizebox{\columnwidth}{!}{%
\begin{tabular}{@{}lcccrlrrlrr@{}}
\toprule
\multicolumn{1}{c}{} & \multicolumn{3}{c}{Party} & \multicolumn{1}{c}{} &  & \multicolumn{2}{c}{\begin{tabular}[c]{@{}c@{}}Coverage Results \\ for 95\% C.I.\end{tabular}} &  & \multicolumn{2}{c}{\begin{tabular}[c]{@{}c@{}}Coverage Results \\ for 99\% C.I.\end{tabular}} \\ \cmidrule(lr){2-4} \cmidrule(lr){7-8} \cmidrule(l){10-11} 
Method & A & B & Others & Avg.Bias &  & Length & No. of dist. &  & Length & No. of dist. \\ \midrule
Truth & 16 (4) & 30 (6) & 1 (0) & - &  & - & - &  & - & - \\
CC & 10 (2) & 35 (7) & 2 (1) & -0.0270 &  & 0.035 & 8 (2) &  & 0.046 & 13 (3) \\
MAR & 11 (2) & 34 (7) & 2 (1) & -0.0183 &  & 0.035 & 17 (4) &  & 0.046 & 28 (7) \\
$\text{FI}_c$ & 18 (6) & 28 (4) & 1 (0) & 0.0073 &  & 0.138 & 42 (9) &  & 0.182 & 45 (10) \\
$\text{FI}_m$ & 14 (5) & 32 (5) & 1 (0) & -0.0067 &  & 0.046 & 21 (8) &  & 0.061 & 31 (10) \\ \bottomrule
\end{tabular}%
}
\end{table}

Table \ref{election prediction} aggregates prediction results across all election districts, presenting the predicted parliamentary seats, average bias, coverage results, the average length of confidence intervals and the number of districts where the confidence interval contains the true value. Due to the small sample sizes in each district and various nonprobability sampling biases, none of the estimators accurately predicted the election results. However, the proposed FI estimators generally outperformed the widely used CC and MAR estimators in practice. Among the FI estimators, the $\text{FI}_{m}$ estimator demonstrates greater accuracy compared to the $\text{FI}_{c}$ estimator, particularly in cases where the nonignorable missing data mechanism model was used. It is also notable that the $\text{FI}_{c}$ estimator provides less stable results compared to the $\text{FI}_{m}$ estimator, due to its reliance on a pre-assumed logistic response model, whereas the $\text{FI}_{m}$ estimator benefits from response model selection. These findings suggest that the  $\text{FI}_{m}$ estimator is a preferable choice over other methods in practical applications.

%==Section 5==%
\section{Discussion}\label{Discussion}
The analysis of nonignorable missing data requires strong assumptions regarding both the response mechanism and the distribution of samples. In this study, we derive sufficient conditions for model identifiability using a generalized linear model and a logistic response mechanism. We then extend the outcome model to accommodate a mixture of distributions belonging to the exponential family and discuss the model identifiability of normal mixture models. {Additionally, we propose an FI-based estimation procedure along with practical guidelines for response model selection.}

{The primary advantage of the proposed FI method is that it obviates the need for identifying an instrumental variable, and the validity of the respondents' outcome model can be assessed solely with observed data.}
Through simulation studies and real data applications, we demonstrated that the FI estimators outperform other methods, including the PPM method. The proposed FI estimators can exhibit lower variability compared to the PPM method, particularly when accounting for the uncertainty in the choice of the sensitivity parameter. Additionally, the FI estimator, when combined with the proposed response model selection procedure, provides relatively robust estimates even in the presence of misspecified response models.

{
However, the proposed FI method has a clear limitation: it also relies on the unverifiable assumption that if the response mechanism is nonignorable, it must be additive, meaning that the missing data \(y\) is related to the response only through the additive term \(\beta y\) \citep{hirano1998combining}. If these assumptions deviate significantly from reality, the proposed method may yield highly biased estimates. On the other hand, in situations where the location of \(y\) is expected to differ substantially between respondents and nonrespondents, this assumption could be reasonably justified.
}

Although we confine the response mechanism to a logistic distribution, it may be possible to utilise other distributions such as the Tobit and Robit models \citep{liu2004robit}. In this case, the integral involved in the observed likelihood cannot be explicitly represented; more unfavourably, this integral may diverge to infinity. Therefore, careful investigation is necessary when employing alternative response mechanisms.

To identify the model, mean structure $\mu(\bm{x};\bm{\gamma})$ must be more complex than $h(\bm{x};\bm{\alpha})$ in Corollary \ref{normal identification}. The model can be easily identified using basis functions, for example the Fourier basis, which is more complex for estimating the mean structure. This model is advantageous because it enhances identifiability and enables a more complex modelling of respondents.

The proposed FI method can be replaced by multiple imputation, which is a popular method of missing data analysis \citep{rubin1978multiple}.
Rubin's variance formula simplifies the calculation of the asymptotic variance of estimators. However, the congeniality condition requires further discussion to guarantee the applicability of Rubin's variance formula, which represents scope of future work. 

\section*{Acknowledgement}
We thank the two referees, the associate editor, and the editor their valuable and constructive comments. The research of Kosuke Morikawa was supported by MEXT Project for Seismology Toward Research Innovation with Data of Earthquake (STAR-E) Grant Number JPJ010217 and JSPS KAKENHI Grant Number JP19186222.
Additionally, research of Jongho Im was supported
by National Research Foundation of Korea (NRF) grants funded by the Korean Government (MSIT) (NRF-2021R1C1C1014407 \& NRF-2022R1A4A1033384).

\bibliographystyle{rss}
\bibliography{glm_mnar.bib}

\appendix
\section*{Supplementary Material}

Due to space constraints, additional theoretical results, simulation results, and further discussion about real data applications are included in the Supplementary Materials. Section 1 contains technical proofs for theorems, lemmas, and corollaries. Section 2 presents additional theorems derived under more flexible conditions. Section 3 discusses variance estimation for the proposed estimator. Section 4 provides further discussion on response model selection from the first simulation study, and Section 5 reports the second simulation study results. For clarity, the equations, figures, tables, and theorems of the supplementary materials are numbered separately from those in the main manuscript.

\section{Technical proofs} \label{Technical proofs}

We first prove Theorem \ref{identify multi}, which is the most general case. Using Theorem \ref{identify multi}, we can prove Theorem \ref{identify} by considering the case for which $K=1$. Theorems \ref{identify} and \ref{identify multi} prove corollaries \ref{normal identification} and \ref{normal_mixture_cor}, respectively.

\begin{proof}[Proof of Theorem \ref{identify multi}]
Using Bayes' theorem, we obtain
\begin{align}
&f(y\mid \bm{x};\bm{\alpha},\beta,\bm{\gamma})P(\delta=1\mid \bm{x},y;\bm{\alpha},\beta)\notag \\
&=\frac{f(y\mid \bm{x}, \delta=1;\bm{\gamma})}{\int f(y\mid \bm{x},\delta=1;\bm{\gamma}) \llp P(\delta=1\mid \bm{x},y;\bm{\alpha},\beta)\rrp^{-1}\ dy  }. \label{change_eq}
\end{align}
When $(\bm{\alpha},\beta,\bm{\gamma})$ and $(\bm{\alpha}^{\prime},\beta^{\prime},\bm{\gamma}^{\prime})$ yield the same observed likelihood, by integrating out $y$ from both sides, we obtain 
\begin{align*}
    \int \frac{f(y\mid \bm{x}, \delta=1;\bm{\gamma}) }{P(\delta=1\mid \bm{x},y;\bm{\alpha},\beta)} dy&=\int \frac{f(y\mid \bm{x}, \delta=1;\bm{\gamma}^{\prime})}{P(\delta=1\mid \bm{x},y;\bm{\alpha}^{\prime},\beta^{\prime})} dy.
\end{align*}
Then, we obtain $f(y\mid \bm{x}, \delta=1;\bm{\gamma})=f(y\mid \bm{x}, \delta=1;\bm{\gamma}^{\prime})$ because both denominators in \eqref{change_eq} are identical.
The identification of $[y\mid \bm{x}, \delta=1]$ reduces our identification problem as
\begin{align*}
\int \frac{f(y\mid \bm{x}, \delta=1;\bm{\gamma}) }{P(\delta=1\mid \bm{x},y;\bm{\alpha},\beta)} dy=\int \frac{f(y\mid \bm{x}, \delta=1;\bm{\gamma})}{P(\delta=1\mid \bm{x},y;\bm{\alpha}^{\prime},\beta^{\prime})} dy \Rightarrow (\bm{\alpha},\beta)=(\bm{\alpha}^{\prime},\beta^{\prime}).
\end{align*}
Next, we show that $\beta=\beta^{\prime}$ is sufficient to show $(\bm{\alpha},\beta)=(\bm{\alpha}^{\prime},\beta^{\prime})$.
Let us introduce a function
\begin{align*}
    l(s)=\int f(y\mid \bm{x},\delta=1;\bm{\gamma})\frac{1}{F(s+\beta y)} dy,
\end{align*}
where $F$ denotes a logistic distribution.
Here, $l(s)$ inherits strict monotonicity from $F(\cdot)$.
When $\beta=\beta^{\prime}$, we obtain $\bm{\alpha}=\bm{\alpha}^{\prime}$ using the following relationship:
\begin{align*}
    l\lp h(\bm{x};\bm{\alpha}) \rp=l\lp h(\bm{x};\bm{\alpha}^{\prime}) \rp \ \Rightarrow \ h(\bm{x};\bm{\alpha})=h(\bm{x};\bm{\alpha}^{\prime})\ \Rightarrow \ \bm{\alpha}=\bm{\alpha}^{\prime}.
\end{align*}
When $f(y\mid \bm{x},\delta=1;\bm{\gamma})$ belongs to a mixture of the exponential family and $P(\delta=1\mid \bm{x},y;\bm{\alpha},\beta)$ is the logistic response mechanism, it can be analogously computed
\begin{align*}
    &\int f(y\mid \bm{x}, \delta=1;\bm{\gamma}) \llp P(\delta=1\mid \bm{x},y;\bm{\alpha},\beta)\rrp^{-1}\ dy\\
    &=\int f(y\mid \bm{x}, \delta=1;\bm{\gamma})
    \llp 1+\exp\lp -h(\bm{x};\bm{\alpha})-\beta y \rp \rrp\ dy\\
    &=1+\exp\lp -h(\bm{x};\bm{\alpha})\rp \int  
    \sum_{k=1}^{K}\pi_k\exp\llp \tau_k \lp y\theta_k -b(\theta_k)\rp +c(y;\tau_k)  \rrp  \cdot \exp\lp -\beta y \rp\ dy\\
    &=1+\exp\lp -h(\bm{x};\bm{\alpha})\rp \sum_{k=1}^{K}\pi_k
    \exp \lp -\tau_k b(\theta_k) \rp \cdot 
    \exp\llp \tau_k b\lp \theta_k-\frac{\beta}{\tau_k} \rp \rrp.
\end{align*}
The identification problem results in
\begin{align*}
    \int \frac{f(y\mid \bm{x}, \delta=1;\bm{\gamma}) }{P(\delta=1\mid \bm{x},y;\bm{\alpha},\beta)} dy&=
    \exp\llp -g(\bm{\alpha},\beta,\bm{\gamma}) \rrp\\
    &=\exp\llp -g(\bm{\alpha}^{\prime},\beta^{\prime},\bm{\gamma}) \rrp
    =\int \frac{f(y\mid \bm{x}, \delta=1;\bm{\gamma})}{P(\delta=1\mid \bm{x},y;\bm{\alpha}^{\prime},\beta^{\prime})} dy,
\end{align*}
where $g(\bm{\alpha},\beta,\bm{\gamma})$ is as defined in Theorem \ref{identify multi}.
Because the assumption $g(\bm{\alpha},\beta,\bm{\gamma})=g(\bm{\alpha}^{\prime},\beta^{\prime},\bm{\gamma}) \Rightarrow \beta=\beta^{\prime}$ guarantees the identification of $\beta$, we obtain the desired identification for $(\bm{\alpha},\beta, \bm{\gamma})$. 
Regarding the necessity, the identifiability of $(\bm{\alpha},\beta)$ can clearly claim the identifiability of $\beta$.
Thus, the theorem is proven.
\end{proof}

\begin{proof}[Proof of Theorem \ref{identify}]
By using Theorem \ref{identify multi}, rearranging the equation $g(\bm{\alpha},\beta,\bm{\gamma})=g(\bm{\alpha}^{\prime},\beta^{\prime},\bm{\gamma})$ provides
\begin{align*}
    h(\bm{x};\bm{\alpha}) - \llp -\tau_b(\theta)+\tau b\lp \theta-\frac{\beta}{\tau} \rp \rrp &=h(\bm{x};\bm{\alpha}^{\prime}) - \llp -\tau_b(\theta)+\tau b\lp \theta-\frac{\beta^{\prime}}{\tau} \rp \rrp \\
    h(\bm{x};\bm{\alpha}) - \tau b\lp \theta-\frac{\beta}{\tau} \rp  &=h(\bm{x};\bm{\alpha}^{\prime}) - \tau b\lp \theta-\frac{\beta^{\prime}}{\tau} \rp .
\end{align*}
Hence, the function above is consistent with \eqref{condition}.
Necessity and sufficiency follow from an argument analogous to the proof of Theorem \ref{identify multi}.
\end{proof}

\begin{proof}[Proof of Theorem \ref{multi Logistic}]
Using the function $g(\bm{\alpha},\beta,\bm{\gamma})$ in Corollary \ref{normal_mixture_cor}, we consider two functions $\exp\{ -g \}$, which are equal but have different parameters, as follows
\begin{align*}
     \sum_{i=1}^{K} \pi_i \exp{\llp-h(\bm{x};\bm{\alpha})-\beta\mu_i(\bm{x};\bm{\kappa}_i)+\frac{1}{2}\beta^2\sigma_i^2\rrp}=
     \sum_{i=1}^{K} \pi_i \exp{\llp -h(\bm{x};\bm{\alpha}^{\prime})-\beta^{\prime}\mu_i(\bm{x};\bm{\kappa}_i)+\frac{1}{2}{\beta^{\prime}}^2\sigma_i^2\rrp}.
\end{align*}
It suffices to show $\beta= \beta^{\prime}$ to prove the identifiability according to Corollary \ref{normal_mixture_cor}.
By employing condition (C2) and Lemma \ref{linear independent}, there exists $K\times K$ permutation matrix $P$ such that 
\begin{align}
   P\beta \mu^{\mathcal{M}}(\bm{x})=\beta^{\prime} \mu^{\mathcal{M}}(\bm{x})\label{per mat eq},
\end{align}
where $\mu^{\mathcal{M}}(\bm{x})=(\mu_1^{\mathcal{M}}(\bm{x};\bm{\kappa}_1),\ldots,\mu_K^{\mathcal{M}}(\bm{x};\bm{\kappa}_K))^{\top}$.
The equation \eqref{per mat eq} leads to
\begin{align*}
    P^n \mu^{\mathcal{M}}(\bm{x})&=P^{n-1}\cdot P \mu^{\mathcal{M}}(\bm{x})\\
    &=P^{n-1}\cdot \frac{\beta^{\prime}}{\beta} \mu^{\mathcal{M}}(\bm{x})=\cdots=\lp \frac{\beta^{\prime}}{\beta}\rp^n \mu^{\mathcal{M}}(\bm{x}).
\end{align*}
Note that since $P$ is the permutation matrix, there exists $n\in\mathbb{N}$ such that $P^n=I$. Thus, there exists $n\in\mathbb{N}$ such that $(\beta^{\prime}/\beta)^n=1$, which implies that $\beta=\beta^{\prime}$ or $\beta=-\beta^{\prime}$.
The Condition (C3) indicates $\beta=\beta^{\prime}$.
When $\beta=-\beta^{\prime}$, equation (\ref{per mat eq}) becomes $ P \mu^{\mathcal{M}}(\bm{x})=- \mu^{\mathcal{M}}(\bm{x}) $, indicating that (C4) is not satisfied. Therefore, this model is identifiable when (C4) holds.
\end{proof}

\begin{proof}[Proof of Theorem \ref{linear and linear 3}]
Following the same approach as in the proof of Theorem \ref{multi Logistic}, we consider the following equation
\begin{align*}
      &  \sum_{i=1}^{K} \pi_i \exp{\llp\lp -\alpha_0-\beta\kappa_{0i}+\frac{1}{2}\beta^2\sigma_i^2\rp-\lp \alpha_1+\beta\kappa_{1i} \rp x\rrp}\\
      & =  \sum_{i=1}^{K} \pi_i \exp{\llp\lp -\alpha_0^{\prime}-\beta^{\prime}\kappa_{0i}+\frac{1}{2}{\beta^{\prime}}^2\sigma_i^2\rp-\lp \alpha_1^{\prime}+\beta^{\prime}\kappa_{1i} \rp x\rrp}.
\end{align*}
It suffices to show $\beta= \beta^{\prime}$ to prove the identifiability according to Theorem \ref{identify multi}.
Using Lemma \ref{linear independent}, there exists a permutation matrix $P$ such that 
\begin{align*}
    P\lp \alpha_1\bm{1}_{K}+\beta \tilde{\bm{\kappa}} \rp=\alpha_1^{\prime}\bm{1}_{K}+\beta^{\prime} \tilde{\bm{\kappa}},
\end{align*}
where $\tilde{\bm{\kappa}}=( \kappa_{11},\ldots,\kappa_{1K} )^{\top}$.
Note that $\bm{1}_{K}$ is an eigenvector of a permutation matrix with an eigenvalue of 1.
Thus, we obtain the following equation
\begin{align}
     P\tilde{\bm{\kappa}}=\frac{(\alpha_1^{\prime}-\alpha_1)}{\beta}\bm{1}_K+ \frac{\beta^{\prime}}{\beta} \tilde{\bm{\kappa}}. \label{key relation}
\end{align}
By applying equation \eqref{key relation} once,
\begin{align*}
    P^2\tilde{\bm{\kappa}}=P\llp \frac{(\alpha_1^{\prime}-\alpha_1)}{\beta} \bm{1}_K+ \frac{\beta^{\prime}}{\beta} \tilde{\bm{\kappa}} \rrp
    =\lp 1+ \frac{\beta^{\prime}}{\beta} \rp\frac{(\alpha_1^{\prime}-\alpha_1)}{\beta} \bm{1}_K+\lp \frac{\beta^{\prime}}{\beta}\rp^2 \tilde{\bm{\kappa}}
\end{align*}
holds, and through repeating this process $n$ times, we get
\begin{align*}
    P^n\tilde{\bm{\kappa}}=\llp 1+\frac{\beta^{\prime}}{\beta}+\cdots+\lp \frac{\beta^{\prime}}{\beta} \rp^{n-1} \rrp\frac{\alpha_1^{\prime}-\alpha_1}{\beta}\bm{1}_K + \lp \frac{\beta^{\prime}}{\beta}\rp^n \tilde{\bm{\kappa}}.
\end{align*}
Note that there exists $n\in\mathbb{N}$ such that $P^n=I$ because $P$ is the permutation matrix.
Thus, the equation $P^n\tilde{\bm{\kappa}}=\tilde{\bm{\kappa}}$ holds for some $n\in\mathbb{N}$ and the following equation is obtained
\begin{align*}
    \llp 1-\lp\frac{\beta^{\prime}}{\beta}\rp^n \rrp\tilde{\bm{\kappa}}=\llp 1+\frac{\beta^{\prime}}{\beta}+\cdots+\lp \frac{\beta^{\prime}}{\beta} \rp^{n-1} \rrp\frac{\alpha_1^{\prime}-\alpha_1}{\beta}\bm{1}_K.
\end{align*}
If $(\beta^{\prime}/\beta)^n\neq 1$ holds, we have $\tilde{\bm{\kappa}}=C\bm{1}_K$, where $C$ is a constant.
However, this result is inconsistent with $\kappa_{1i}\neq\kappa_{1j}\ (i\neq j)$.
Hence, we obtain $(\beta^{\prime}/\beta)^n= 1$, meaning that $\beta=\beta^{\prime}$ or $\beta=-\beta^{\prime}$.
Under the first condition of Theorem \ref{linear and linear 3}, $\beta=\beta^{\prime}$ is immediately apparent.

Next, we show that $\beta=\beta^{\prime}$ under the second condition of Theorem \ref{linear and linear 3}. Based on the above argument, if we assume $\beta=-\beta^{\prime}$, equation \eqref{key relation} provides
\begin{align*}
  P\tilde{\bm{\kappa}}
  =\frac{(\alpha_1^{\prime}-\alpha_1)}{\beta}\bm{1}_K-\tilde{\bm{\kappa}},
\end{align*}
which contradicts the second condition of Theorem \ref{linear and linear 3}. Therefore, we obtain $\beta=\beta^{\prime}$.
\end{proof}

\begin{proof}[Proof of Theorem \ref{linear and linear}]
Following the same approach as in the proof of Theorem \ref{multi Logistic}, we consider the following equation
\begin{align*}
      & \sum_{i=1}^{2} \pi_i \exp{\llp\lp -\alpha_0-\beta\kappa_{0i}+\frac{1}{2}\beta^2\sigma_i^2\rp-\lp \alpha_1+\beta\kappa_{1i} \rp x\rrp};\\
      & =  \sum_{i=1}^{2} \pi_i \exp{\llp\lp -\alpha_0^{\prime}-\beta^{\prime}\kappa_{0i}+\frac{1}{2}{\beta^{\prime}}^2\sigma_i^2\rp-\lp \alpha_1^{\prime}+\beta^{\prime}\kappa_{1i} \rp x\rrp}.
\end{align*}
It suffices to show $\beta= \beta^{\prime}$ to prove the identifiability according to Theorem \ref{identify multi}.
Using Lemma \ref{linear independent}, one of the following equations holds: 
\begin{align*}
    &\mathrm{Case \ 1:} \quad \alpha_1+\beta\kappa_{11}=\alpha_1^{\prime}+\beta^{\prime}\kappa_{11}, \quad \alpha_1+\beta\kappa_{12}=\alpha_1^{\prime}+\beta^{\prime}\kappa_{12};\\
    &\mathrm{Case \ 2:} \quad \alpha_1+\beta\kappa_{11}=\alpha_1^{\prime}+\beta^{\prime}\kappa_{12}, \quad
    \alpha_1+\beta\kappa_{12}=\alpha_1^{\prime}+\beta^{\prime}\kappa_{11}.
\end{align*}

Under Case 1, subtracting both equations gives $\beta(\kappa_{11}-\kappa_{12})=\beta^{\prime}(\kappa_{11} - \kappa_{12})$.
Therefore, we obtain $\beta=\beta^{\prime}$ from the assumption $\kappa_{11}\neq \kappa_{12}$.

Under Case 2, a similar calculation of Case 1 yields $\beta=-\beta^{\prime}$.
Next, we compare the constant part. If the two models are not identifiable, following two equations hold 
\begin{align*}
    -\alpha_0-\beta\kappa_{01}+\frac{1}{2}\beta^2\sigma_1^2+\log\pi_1&=-\alpha_0^{\prime}-\beta^{\prime}\kappa_{02}+\frac{1}{2}{\beta^{\prime}}^2\sigma_2^2+\log\pi_2,\\
    -\alpha_0-\beta\kappa_{02}+\frac{1}{2}\beta^2\sigma_2^2+\log\pi_2&=-\alpha_0^{\prime}-\beta^{\prime}\kappa_{01}+\frac{1}{2}{\beta^{\prime}}^2\sigma_1^2+\log\pi_1.
\end{align*}
Because $\beta=-\beta^{\prime}$, rearranging the equation above leads to
\begin{align*}
    \beta^2(\sigma_1^2-\sigma_2^2)= 2\lp \log\pi_2-\log\pi_1 \rp.
\end{align*}
The above equation is inconsistent with conditions 2 and 3 of Theorem \ref{linear and linear}.
\end{proof}
The following lemma shows the linear independence of exponentials of multivariate polynomials. This result plays an important role in deriving the identification conditions for normal mixtures.
A related proof exists on the Stack Exchange website, and we provide it here in a more extended form.
\begin{lemma}\label{linear independent}
Let $\bm{x}$ be the $p$-dimensional vector $(x_1,\ldots,x_p)^{\top}$, $P_1(\bm{x}),\ldots,P_n(\bm{x})$ be distinct multivariate polynomials without constant term, $R_1(\bm{x}),\ldots,R_n(\bm{x})$ be rational functions of multivariate polynomials, and a domain of all these functions be the subset of the Euclidean space $\mathbb{R}^p$ which contains an interior point. Then, the following result holds:
\begin{align*}
    R_1(\bm{x})e^{P_1(\bm{x})} + \cdots + R_n(\bm{x})e^{P_n(\bm{x})} = 0 \ \Rightarrow \ R_1(\bm{x})=\cdots=R_n(\bm{x})=0.
\end{align*}
\end{lemma}

Note that Lemma \ref{linear independent} implies the linear independence of exponentials of multivariate polynomials when the functions $R_1(\bm{x}),\ldots,R_n(\bm{x})$ are constant.
\begin{proof}[Proof of Lemma \ref{linear independent}]
We prove this through mathematical induction.
The case for $n=1$ is straightforward.
Now, we assume that the case for $n=k-1$ is true. Let 
\begin{align*}
    R_1(\bm{x})e^{P_1(\bm{x})} + \cdots + R_k(\bm{x})e^{P_k(\bm{x})} = 0,
\end{align*}
where $P_1(\bm{x}),\ldots,P_k(\bm{x})$ are distinct polynomials without a constant term and $R_1(\bm{x}),\ldots,R_k(\bm{x})$ are rational functions. If $R_k(\bm{x})\neq0$, it follows 
\begin{align}
     \frac{R_1(\bm{x})}{R_k(\bm{x})}e^{P_1(\bm{x})-P_k(\bm{x})} + \cdots +  \frac{R_{k-1}(\bm{x})}{R_k(\bm{x})}e^{P_{k-1}(\bm{x})-P_k(\bm{x})}+1  = 0. \label{rational}
\end{align}
Since both sides are differentiable at the interior point of the domain of all functions, differentiating of equation \eqref{rational} with respect to $x_l$ gives
\begin{align*}
     \sum_{i=1}^{k-1}\lllp \frac{d}{dx_l}\lp \frac{R_i(\bm{x})}{R_k(\bm{x})} \rp + \frac{R_i(\bm{x})}{R_k(\bm{x})}\cdot\frac{d}{dx_l}\llp P_i(\bm{x})-P_k(\bm{x}) \rrp  \rrrp e^{P_i(\bm{x})-P_k(\bm{x})}=0.
\end{align*}
Because $P_1-P_k,\ldots, P_{k-1}-P_k$ are distinct multivariate polynomials without a constant term, the assumption of $n=k-1$ case yields
\begin{align*}
     \frac{d}{dx_l}\lp \frac{R_i(\bm{x})}{R_k(\bm{x})} \rp + \frac{R_i(\bm{x})}{R_k(\bm{x})}\cdot\frac{d}{dx_l}\lp P_i(\bm{x})-P_k(\bm{x}) \rp  &=0,\\
    \llp \frac{d}{dx_l}\lp \frac{R_i(\bm{x})}{R_k(\bm{x})} \rp + \frac{R_i(\bm{x})}{R_k(\bm{x})}\cdot\frac{d}{dx_l}\lp P_i(\bm{x})-P_k(\bm{x}) \rp  \rrp e^{P_i(\bm{x})-P_k(\bm{x})}&=0,
\end{align*}
where $i=1,2,\ldots,k-1$. 
Integrating out $x_l$ from both sides, we obtain
\begin{align*}
    \frac{R_i(\bm{x})}{R_k(\bm{x})}e^{P_i(\bm{x})-P_k(\bm{x})}=C_i(x_1,\ldots,x_{l-1},x_{l+1},\ldots,x_p),
\end{align*}
where $l=1,2,\cdots,p$.
The left-hand side is constant because it does not depend on $l$. Therefore, it is denoted by $C_i$. 
If $C_i\neq 0$, $R_i(\bm{x})/R_k(\bm{x})$ and $P_i(\bm{x})-P_k(\bm{x})$ are constants, which contradicts the fact that $P_1(\bm{x}),\ldots,P_n(\bm{x})$ are distinct polynomials without constant terms.
Thus, using $C_i= 0$ for $i=1,2,\ldots, k-1$ and the equation \eqref{rational}, the following contradictory equation holds
\begin{align*}
    0=\frac{R_1(\bm{x})}{R_k(\bm{x})}e^{P_1(\bm{x})-P_k(\bm{x})} + \cdots +  \frac{R_{k-1}(\bm{x})}{R_k(\bm{x})}e^{P_{k-1}(\bm{x})-P_k(\bm{x})}+1 =1.
\end{align*}
Consequently, we obtain $R_k(\bm{x})=0$ that follows
\begin{align*}
    R_1(\bm{x})e^{P_1(\bm{x})} + \cdots + R_{k-1}(\bm{x})e^{P_{k-1}(\bm{x})} = 0.
\end{align*}
From the assumption of $n=k-1$, $R_1(\bm{x})=\cdots=R_{k-1}(\bm{x})=0$ holds and the lemma is proven.
\end{proof}

\section{Further Theorems}
Although Example \ref{single gen} and Theorem \ref{multi Logistic} indicate the importance of condition (C2), eliminating this condition enables a more flexible model. Thus, we derive sufficient conditions for the identifiability of a mixture of simple linear regression models:
\begin{align}
    &[y\mid \bm{x}, \delta=1;\bm{\gamma}]\sim \sum_{k=1}^{K}\pi_k N(\kappa_{0k}+\kappa_{1k}x,\sigma_k^2) \label{simple normal mixture reg},  
\end{align}
and $\mathcal{H}=\{ 1,x \}$, which do not satisfy (C2). Because the sufficient conditions for model identifiability differ for $K\geq 3$ and $K=2$, we derive the conditions separately.

\begin{theorem}\label{linear and linear 3}
Suppose that the response mechanism is (\ref{Resp}) with $h(\bm{x}; \bm{\alpha})=\alpha_0 +\alpha_1 x$ and the distribution of $[y\mid \bm{x},\delta=1]$ is identifiable and has a normal mixture density in (\ref{simple normal mixture reg}) with the number of mixture components $K\geq 3$. We further assume that $\mathcal{H}=\{1, x \}$, $\tilde{\bm{\kappa}}=(\kappa_{11},\kappa_{12},\ldots,\kappa_{1K})^{\top}$ is a vector of first-order coefficients, and $\kappa_{1i}\neq \kappa_{1j}\ (i\neq j)$. The model is identifiable if at least one of the following conditions is satisfied:
\begin{itemize}
    \item[1] Sign of $\beta$ is known;
    \item[2] For all $K\times K$ permutation matrices $P$ and for all $r\in\mathbb{R}$, 
    \begin{align*}
        \lp P+I \rp\tilde{\bm{\kappa}}  \neq r\bm{1}_{K},
    \end{align*}
    where $\bm{1}_{n}$ denotes an $n\times1$ vector of ones.
\end{itemize}
\end{theorem}

To clarify the second condition, we consider $K=3$ and assume that $\kappa_{11}>\kappa_{12}>\kappa_{13}$ without loss of generality. If $P$ is defined as 
\begin{align*}
    P=\begin{pmatrix}
0 & 0 & 1\\
0 & 1 & 0\\
1 & 0 & 0\\
\end{pmatrix}
,
\end{align*}
we obtain $ (P+I)\tilde{\bm{\kappa}}=(\kappa_{11}+\kappa_{13}, 2\kappa_{12}, \kappa_{11}+\kappa_{13})^{\top}$.
Then, the second condition is satisfied, unless $2\kappa_{12}=\kappa_{11}+\kappa_{13}$. More importantly, it does not generally hold and can be tested using the observed data.

However, for $K=2$, the second condition in Theorem \ref{linear and linear 3} does not hold for any model in $[y\mid \bm{x},\delta=1]$, which can be demonstrated by assuming $\kappa_{11}>\kappa_{12}$ without loss of generality. Using a similar argument, we derive $(P+I)\tilde{\bm{\kappa}}=(\kappa_{11}+\kappa_{12}, \kappa_{11}+\kappa_{12})^{\top}=r\bm{1}_{2}$, where
\begin{align*}
     P=\begin{pmatrix}
 0 & 1\\
 1 & 0\\
\end{pmatrix}
, \ r=\kappa_{11}+\kappa_{12}.
\end{align*}
Thus, the second condition does not hold for any model $[y\mid \bm{x},\delta=1]$. Therefore, a more careful investigation is required for $K=2$. 

\begin{theorem}\label{linear and linear}
Suppose that the response mechanism is (\ref{Resp}) and the distribution of $[y\mid x,\delta=1]$ is identifiable and has a normal mixture density in (\ref{simple normal mixture reg}) with the number of mixture components $K=2$. We further assume that $\mathcal{H}=\{1, x \}$, $\mu_1(x,\bm{\kappa}_1)=\kappa_{01}+\kappa_{11}x\ (\kappa_{11}\neq0),\ \mu_2(x,\bm{\kappa}_2)=\kappa_{02}+\kappa_{12}x\ (\kappa_{12}\neq0),$ and $h(\bm{x};\bm{\alpha})=\alpha_0+\alpha_1x \ (\alpha_1\neq0)$. The model is then identifiable if the following conditions hold:
\begin{itemize}
    \item[1] $\kappa_{11}\neq\kappa_{12}$;
    \item[2] $\sigma_1=\sigma_2\Rightarrow \pi_1\neq\pi_2$;
    \item[3] $\sigma_1\neq\sigma_2\Rightarrow \ \lp \log\pi_2-\log\pi_1 \rp  \lp \sigma_1^2-\sigma_2^2\rp^{-1}\leq0$.
\end{itemize}
\end{theorem}
Overall, the conditions required in Theorems \ref{linear and linear} are more difficult to satisfy than those in Theorems \ref{linear and linear 3}. Although these conditions can be verified using the observed data, they are redundant.

\section{Variance estimation} \label{Variance estimation}
Under some regularity conditions, \cite{riddles2016propensity} proved that the estimator $\hat{\bm{\phi}}$ obtained by FI has asymptotic normality:
\begin{align*}
    \sqrt{n}\lp \hat{\bm{\phi}} - \bm{\phi}_0 \rp\to N(0,\Sigma_{\bm{\phi}}),
\end{align*}
where 
\begin{align*}
    \Sigma_{\bm{\phi}}&=\mathcal{I}_{22}^{-1}\mathrm{var}\llp s_2(\bm{\phi}_0;\bm{\gamma}_0)- \mathcal{I}_{21}\mathcal{I}_{11}^{-1}s_1(\bm{\gamma}_0) \rrp \lp\mathcal{I}_{22}^{-1}\rp^{\top},\\
    \mathcal{I}_{11}&=\mathrm{var}\llp s_1(\bm{\gamma}_0) \rrp,\\
    \mathcal{I}_{21}&=-E\lllp (1-\delta) \llp s(\bm{\phi}_0) -\bar{s}_0(\bm{\phi}_0;\bm{\gamma}_0) \rrp s_1^{\top}(\bm{\gamma}_0) \rrrp,\\
    \mathcal{I}_{22}&=-E\lllp (1-\delta) \bar{s}_0(\bm{\phi}_0;\bm{\gamma}_0)\bm{\bar{z}}_0^{\top}(\bm{x};\bm{\phi}_0),  \rrrp
\end{align*}
$s_1(\bm{\gamma})=\lp \partial/ \partial \bm{\gamma} \rp\delta \log f(y\mid \bm{x},\delta=1;\bm{\gamma})$,
$s_2(\bm{\phi};\bm{\gamma})=\delta s(\bm{\phi})+(1-\delta) \bar{s}_0(\bm{\phi};\bm{\gamma})$,
$\bar{s}_0(\bm{\phi};\bm{\gamma})=E\llp s(\bm{\phi}) \mid \bm{x},\delta=0;\bm{\gamma} \rrp$,
$\bar{\bm{z}}_0(\bm{x};\bm{\phi})=E\lllp \bm{z}(\bm{x},y;\bm{\alpha},\beta) \mid \bm{x},\delta=0 \rrp$
and $s(\bm{\phi})=S(\bm{\alpha},\beta;\bm{x},y,\delta)$ is defined in (\ref{mean score}). Therefore, the asymptotic variance of $\hat{\bm{\phi}}$ can be estimated by
\begin{align*}
     \frac{1}{n}\hat{\mathcal{I}}_{22}^{-1}\llp 
      \sum_{i:\delta_i=1}  \bm{J}_i \bm{J}_i^{\top} + \sum_{i:\delta_i=0}  \bar{s}_0^*\lp \hat{\bm{\phi}};\bm{x}_i\rp  {\bar{s}_0}^{*\top}\lp \hat{\bm{\phi}};\bm{x}_i\rp   \rrp  \lp\hat{\mathcal{I}}_{22}^{-1}\rp^{\top},
\end{align*}
where
\begin{align*}
    \bm{J}_i&= S(\hat{\bm{\phi}};\bm{x}_i,y_i,\delta_i) -\hat{\mathcal{I}}_{21}\hat{\mathcal{I}}_{11}^{-1} \left. \frac{\partial}{\partial \bm{\gamma}}  \log f(y_i\mid \bm{x}_i,\delta=1;\bm{\gamma})\right|_{\bm{\gamma}=\hat{\bm{\gamma}}},\\
    \bar{s}_0^*\lp \hat{\bm{\phi}};\bm{x}_i\rp &= \sum_{j=1}^{M}   w_{ij}^*(\hat{\bm{\phi}}) S(\hat{\bm{\phi}};\bm{x}_i,y_i^{*(j)},\delta_i),\\ 
    \hat{\mathcal{I}}_{11}&= \frac{1}{n} \sum_{i:\delta_i=1} \llp\bm{s}_1(\hat{\bm{\gamma}}:\bm{x}_i,y_i) \rrp\llp\bm{s}_1(\hat{\bm{\gamma}}:\bm{x}_i,y_i) \rrp^{\top} ,  \\
    \hat{\mathcal{I}}_{21}&= \sum_{i=1}^{n} (1-\delta_i) \sum_{j=1}^{M} w_{ij}^*(\hat{\bm{\phi}}) \llp  S(\hat{\bm{\phi}};\bm{x}_i,y_i^{*(j)},\delta_i)- \bar{s}_0^*\lp \hat{\bm{\phi}};\bm{x}_i\rp \rrp \bm{s}_1^{\top}(\hat{\bm{\gamma}};\bm{x}_i,y_i^{*(j)}),\\
     \hat{\mathcal{I}}_{22}&=\sum_{i=1}^{n} (1-\delta_i)\bar{s}_0^*\lp \hat{\bm{\phi}};\bm{x}_i\rp \sum_{j=1}^{M} w_{ij}^*(\hat{\bm{\phi}}) \bm{z}^{\top}(\bm{x}_i,y_i^{*(j)};\hat{\bm{\phi}}) ,\\
     \bm{s}_1(\bm{\gamma}:\bm{x},y)&=\frac{\partial}{\partial \bm{\gamma}}  \log f(y\mid \bm{x},\delta=1;\bm{\gamma}),\\
     w_{ij}^*(\bm{\phi})&=\frac{\exp{(-\beta y_i^{*(j)})}}{\sum_{k=1}^{M} \exp{(-\beta y_i^{*(k)})}},
\end{align*}
and \(y_i^{*(j)}\) (\(j=1, \dots, M\)) is a random sample generated from \(f(y \mid \boldsymbol{x}_i, \delta_i=1; \hat{\bm{\gamma}})\).
\cite{riddles2016propensity} provided that $\mathcal{I}_{21}=\bm{0}$ and $\mathcal{I}_{22}=\mathrm{var}\{ s(\bm{\phi}) \}$ under MAR assumption.
In this case, $\Sigma_{\bm{\phi}}=\mathcal{I}_{22}^{-1}$, which is consistent with the asymptotic variance of the maximum likelihood estimator of the response model, and Section 4 of \cite{riddles2016propensity} also gave the asymptotic variance of the IPW estimator of the mean, which is consistent with the asymptotic variance of the IPW estimator using the MAR working model.

\section{Further discussion on Simulation Study I}

As shown in Tables \ref{sim1_point} and \ref{var_sim1}, the proposed FI estimator performs poorly in nonidentifiable scenarios. To address this issue, we have proposed a practical guideline in Section 3.3 for selecting a final response model.

\begin{figure}[h]
\caption{\label{pval_boxplot}Box plots of p-values for testing the null hypothesis $\beta=0$, categorized by the ratio $\theta_2/\theta_1$.}
\centering
\includegraphics[width=\columnwidth]{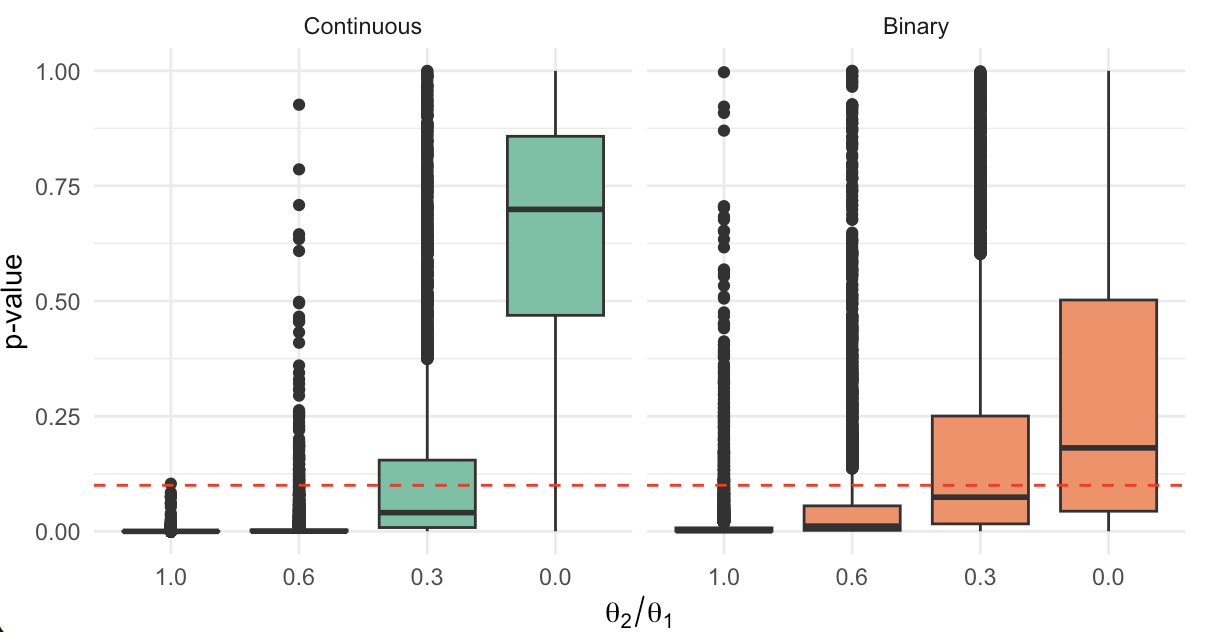}
\end{figure}

In the first simulation study, specifically, we utilize p-values for the null hypothesis \(\beta = 0\) as an empirical measure of a model's degree of identifiability. Figure \ref{pval_boxplot} presents box plots of the p-values across different \(\theta_2/\theta_1\) ratios for each outcome case, with the red dashed line indicating the significance level of 0.1. A noticeable trend of increasing p-values is observed as identifiability weakens (i.e., as \(\theta_2/\theta_1\) decreases).  In the continuous case, the nonidentifiable model with \(\theta_2/\theta_1 = 0\) fails to reject the null hypothesis in the majority of Monte Carlo samples. These findings support the approach of substituting the challenge of determining model identifiability with a hypothesis test of whether $\beta$ equals to 0.

Naturally, there are instances where the model is identifiable, but the null hypothesis cannot be rejected because the missing mechanism is nearly ignorable. In such cases, the conclusion remains the same: it is preferable to choose an ignorable missing data mechanism model rather than a nonignorable one.

\section{Second Simulation Study}
In the second simulation study, we assess the robustness of our proposed method against model misspecification, comparing it with the PPM approach introduced by \cite{andridge2011proxy}. Specifically, we compare the maximum likelihood estimators of the normal PPM model across sensitivity parameters \(\lambda \in \{0, 1, \infty\}\). The fitted value of \(y\), regressed on \(x_1\) and \(x_2\), is used as the proxy variable. For implementation, we utilized the functions provided by \cite{andridge2020proxy}, available on their GitHub at \href{http://github.com/randridge/PPMA}{http://github.com/randridge/PPMA}.

To simulate scenarios of outcome model misspecification, we generate \((x_1, x_2, y)\) from the following outcome models: 

For \(x_1 \sim N(0.5, 0.5)\), \(x_2 \sim N(0.5, 0.5)\), and \(y = \theta_0 + \theta_1 x_1 + \theta_2 x_2 + \epsilon\):

\begin{itemize}
    \item Normal (baseline): \(\epsilon \sim N(0,\; \sigma^2)\)
    \item Normal with heterogeneous variance: \(\epsilon \sim N(0,\; \psi(1 + x_1^2))\)
    \item Log-normal: \(\epsilon \sim \text{Lognormal}(-\xi/2,\; \xi)\)
    \item Gamma: \(x_1 \sim \text{Gamma}(0.5, 1)\), \(x_2 \sim \text{Gamma}(0.5, 1)\), and \(\epsilon \sim N(0, \sigma^2)\). The gamma distributions are characterized by the shape and scale parameters.
\end{itemize}

We set the outcome model parameters \((\theta_0, \theta_1, \theta_2, \sigma, \xi, \psi)\) so that \(\theta_2/\theta_1 = 1\), \(E(y) = \text{Var}(y) = 0.5\), and \(\rho\) assumes value of 0.5 (weak proxy) or 0.8 (strong proxy). Since the true outcome distribution may be non-normal, the FI estimators semi-parametrically estimate the respondents' outcome model by positing a normal mixture model:
\[
[y \mid \boldsymbol{x}, \delta=1; \boldsymbol{\gamma}] \sim \sum_{k=1}^{2} \pi_k N(\kappa_{0k} + \kappa_{1k} x_1 + \kappa_{2k} x_2, \sigma_k^2).
\]

The response indicator \(\delta\) is generated from a logistic regression model: \(\text{logit}\{P(\delta=1 \mid \boldsymbol{x}, y)\} = \alpha_0 + \alpha_1 x_1 + \beta y\). The values of \((\alpha_1, \beta)\) are set according to the type of response mechanism: \((\alpha_1, \beta) = (2, 0)\) for ignorable missing data mechanism model ($\text{RM}_0$), \((\alpha_1, \beta) = (1, 1)\) for the first nonignorable missing mechanim ($\text{RM}_1$), and \((\alpha_1, \beta) = (0, 2)\) for the second nonignorable missing mechanim ($\text{RM}_{2}$). The parameter \(\alpha_0\) is adjusted to ensure that the response rate is approximately 70\%.

For the cases of response model misspecification, we first fix the outcome model as the baseline normal case and generate \(\delta\) from a binomial response model with \(P(\delta=1 \mid \boldsymbol{x}, y) = g(\alpha_0 + \alpha_1 x_1 + \beta y)\), where \(g(\cdot)\) is one of the following link functions:
\begin{itemize}
    \item Logistic (baseline): \(g(w) = \frac{1}{1 + \exp(-w)}\)
    \item Probit: \(g(w) = \Phi(w)\)
    \item Log-log: \(g(w) = \exp(-\exp(-w))\)
    \item Complementary log-log: \(g(w) = 1 - \exp(-\exp(w))\).
\end{itemize}

The values of the outcome and response model parameters are set similarly to those used in the simulation for outcome model misspecification. For each scenario, we generate \(B = 1{,}000\) independent Monte Carlo samples, each with a sample size of \(n = 1{,}000\). The imputation size $M$ for the FI estimators is set to 100.

We set \(\text{FI}_c\) as follows: for response mechanism \(\text{RM}_1\), we use \(\text{logit}\{P(\delta=1 \mid \boldsymbol{x}, y)\} = \alpha_0 + \alpha_1 x_1 + \beta y\); for response mechanism \(\text{RM}_2\), we use \(\text{logit}\{P(\delta=1 \mid \boldsymbol{x}, y)\} = \alpha_0 + \beta y\). In the case of response mechanism of \(\text{RM}_{0}\), where the MAR estimator may posit \(\text{logit}\{P(\delta=1 \mid \boldsymbol{x})\} = \alpha_0 + \alpha_1 x_1\), \(\text{FI}_c\) is set to use \(\text{logit}\{P(\delta=1 \mid \boldsymbol{x}, y)\} = \alpha_0 + \alpha_1 x_1 + \beta y\). This approach avoids redundant duplication and allows us to assess the efficiency of the fractional imputation estimator as \(\beta\) shrinks to zero.

Table \ref{sim2_outcome} presents the Monte Carlo means of RB and RMSE for each method in estimating \(E(y)\) across various combinations of \(\rho\), response mechanisms, and outcome models. All $\text{PPM}_{\lambda}$ estimators show significant bias when their underlying assumptions do not align with the simulation conditions. For example, the estimates from $\text{PPM}_0$ estimator are biased under the nonignorable missing mechanisms $\text{RM}_1$ and $\text{RM}_2$, while $\text{PPM}_{\infty}$ exhibits substantial bias unless the missing mechanism is $\text{RM}_2$. Additionally, the biases in the PPM estimators are particularly pronounced for non-normal outcome models. Unlike the FI estimators, the \(\text{PPM}_{\lambda}\) 
estimators assume that the proxy variable and $y$ follow a bivariate normal distribution given \(\delta\). As a result, the \(\text{PPM}_{\lambda}\) estimators are especially vulnerable to errors when the distributions of \(x_1\), \(x_2\), and \(y \mid (x_1, x_2)\) deviate significantly from normality. 
In contrast, both fractional imputation estimators produce nearly unbiased estimates in all scenarios. Notably,
the \(\text{FI}_c\) estimator maintains its unbiasedness even when the response model is mis-specified, such as in the case of $\text{RM}_0$. 

\begin{table}
\caption{\label{sim2_outcome}
Monte Carlo means of Relative Bias (RB, \%) and Root Mean Square Error (RMSE) of the point estimators for $E(y)$ in cases where the outcome model is misspecified.}
\centering
\resizebox{\textwidth}{!}{%
\begin{tabular}{@{}cllrrlrrlrrlrrlrrlrr@{}}
\toprule
 &  &  & \multicolumn{8}{c}{$\rho=0.5$} &  & \multicolumn{8}{c}{$\rho=0.8$} \\ \cmidrule(lr){4-11} \cmidrule(l){13-20} 
 & \multicolumn{1}{c}{} &  & \multicolumn{2}{c}{$\text{RM}_{0}$} &  & \multicolumn{2}{c}{$\text{RM}_1$} &  & \multicolumn{2}{c}{$\text{RM}_{2}$} &  & \multicolumn{2}{c}{$\text{RM}_{0}$} &  & \multicolumn{2}{c}{$\text{RM}_1$} &  & \multicolumn{2}{c}{$\text{RM}_{2}$} \\ \cmidrule(lr){4-5} \cmidrule(lr){7-8} \cmidrule(lr){10-11} \cmidrule(lr){13-14} \cmidrule(lr){16-17} \cmidrule(l){19-20} 
Outcome model & Method &  & RB & RMSE &  & RB & RMSE &  & RB & RMSE &  & RB & RMSE &  & RB & RMSE &  & RB & RMSE \\ \midrule
\multirow{7}{*}{\begin{tabular}[c]{@{}c@{}}Normal\\ (baseline)\end{tabular}} & CC &  & 17.1 & 0.090 &  & 33.9 & 0.171 &  & 48.2 & 0.242 &  & 27.3 & 0.139 &  & 38.8 & 0.195 &  & 48.3 & 0.242 \\
 & MAR &  & -0.1 & 0.034 &  & 22.3 & 0.115 &  & 38.4 & 0.193 &  & -0.1 & 0.034 &  & 11.0 & 0.061 &  & 20.6 & 0.106 \\
 & $\text{PPM}_{0}$ &  & 0.0 & 0.028 &  & 21.4 & 0.110 &  & 38.3 & 0.193 &  & 0.0 & 0.025 &  & 10.7 & 0.059 &  & 20.3 & 0.104 \\
 & $\text{PPM}_{1}$ &  & -18.6 & 0.098 &  & 5.8 & 0.041 &  & 26.4 & 0.134 &  & -7.6 & 0.046 &  & 2.2 & 0.027 &  & 11.6 & 0.063 \\
 & $\text{PPM}_{\infty}$ &  & -57.8 & 0.294 &  & -29.7 & 0.156 &  & -0.2 & 0.042 &  & -17.3 & 0.091 &  & -9.0 & 0.053 &  & 0.1 & 0.027 \\
 & $\text{FI}_{c}$ &  & -0.2 & 0.051 &  & 0.8 & 0.044 &  & 3.7 & 0.046 &  & -0.1 & 0.035 &  & 0.1 & 0.033 &  & 1.5 & 0.036 \\
 & $\text{FI}_{m}$ &  & -0.2 & 0.043 &  & 1.6 & 0.050 &  & 4.0 & 0.048 &  & -4.3 & 0.073 &  & -3.3 & 0.051 &  & 1.5 & 0.036 \\ \midrule
\multirow{7}{*}{\begin{tabular}[c]{@{}c@{}}Normal\\ (heterogeneous\\ variance)\end{tabular}} & CC &  & 17.0 & 0.090 &  & 30.2 & 0.153 &  & 43.6 & 0.219 &  & 27.3 & 0.139 &  & 37.1 & 0.187 &  & 46.1 & 0.232 \\
 & MAR &  & -0.2 & 0.033 &  & 17.8 & 0.093 &  & 33.4 & 0.168 &  & -0.1 & 0.034 &  & 8.8 & 0.052 &  & 18.0 & 0.094 \\
 & $\text{PPM}_{0}$ &  & 0.0 & 0.025 &  & 16.4 & 0.085 &  & 32.9 & 0.166 &  & 0.0 & 0.024 &  & 8.2 & 0.047 &  & 17.3 & 0.089 \\
 & $\text{PPM}_{1}$ &  & -18.9 & 0.099 &  & 0.6 & 0.026 &  & 22.7 & 0.116 &  & -8.1 & 0.047 &  & -0.7 & 0.025 &  & 8.6 & 0.050 \\
 & $\text{PPM}_{\infty}$ &  & -59.4 & 0.302 &  & -33.7 & 0.175 &  & 2.6 & 0.040 &  & -18.5 & 0.097 &  & -12.5 & 0.068 &  & -2.6 & 0.030 \\
 & $\text{FI}_{c}$ &  & -0.1 & 0.056 &  & -3.8 & 0.049 &  & 2.3 & 0.044 &  & -0.1 & 0.034 &  & -0.6 & 0.033 &  & 1.5 & 0.035 \\
 & $\text{FI}_{m}$ &  & -0.3 & 0.045 &  & -2.6 & 0.050 &  & 3.1 & 0.050 &  & -1.9 & 0.052 &  & -2.8 & 0.063 &  & 1.6 & 0.036 \\ \midrule
\multirow{7}{*}{Log-normal} & CC &  & 17.0 & 0.089 &  & 30.9 & 0.157 &  & 40.5 & 0.204 &  & 27.3 & 0.139 &  & 38.2 & 0.192 &  & 46.7 & 0.234 \\
 & MAR &  & -0.1 & 0.034 &  & 19.5 & 0.102 &  & 29.9 & 0.152 &  & -0.1 & 0.034 &  & 10.5 & 0.060 &  & 18.7 & 0.097 \\
 & $\text{PPM}_{0}$ &  & 0.1 & 0.028 &  & 18.4 & 0.096 &  & 29.8 & 0.152 &  & 0.0 & 0.025 &  & 10.1 & 0.057 &  & 18.3 & 0.095 \\
 & $\text{PPM}_{1}$ &  & -18.5 & 0.097 &  & 0.1 & 0.029 &  & 12.8 & 0.070 &  & -7.5 & 0.045 &  & 0.6 & 0.025 &  & 7.5 & 0.045 \\
 & $\text{PPM}_{\infty}$ &  & -57.8 & 0.294 &  & -45.4 & 0.234 &  & -31.7 & 0.168 &  & -17.3 & 0.090 &  & -12.2 & 0.066 &  & -7.3 & 0.046 \\
 & $\text{FI}_{c}$ &  & 1.2 & 0.090 &  & -4.3 & 0.041 &  & -2.6 & 0.032 &  & -0.1 & 0.035 &  & -0.5 & 0.031 &  & -0.2 & 0.033 \\
 & $\text{FI}_{m}$ &  & -0.7 & 0.041 &  & -2.6 & 0.049 &  & -1.5 & 0.042 &  & -2.1 & 0.046 &  & -1.5 & 0.035 &  & -0.2 & 0.033 \\ \midrule
\multirow{7}{*}{Gamma} & CC &  & 10.5 & 0.060 &  & 31.2 & 0.158 &  & 46.8 & 0.235 &  & 16.8 & 0.089 &  & 31.4 & 0.159 &  & 42.0 & 0.212 \\
 & MAR &  & 0.1 & 0.027 &  & 22.4 & 0.115 &  & 38.4 & 0.193 &  & 0.1 & 0.026 &  & 11.1 & 0.060 &  & 20.7 & 0.106 \\
 & $\text{PPM}_{0}$ &  & 0.1 & 0.027 &  & 21.5 & 0.110 &  & 38.3 & 0.193 &  & 0.1 & 0.025 &  & 10.8 & 0.059 &  & 20.4 & 0.104 \\
 & $\text{PPM}_{1}$ &  & -9.3 & 0.054 &  & 12.1 & 0.066 &  & 30.2 & 0.153 &  & -3.6 & 0.031 &  & 6.3 & 0.040 &  & 15.7 & 0.082 \\
 & $\text{PPM}_{\infty}$ &  & -27.2 & 0.140 &  & -6.7 & 0.047 &  & 14.4 & 0.078 &  & -8.1 & 0.048 &  & 0.7 & 0.026 &  & 9.9 & 0.055 \\
 & $\text{FI}_{c}$ &  & -0.3 & 0.047 &  & 0.2 & 0.047 &  & 3.8 & 0.049 &  & 0.0 & 0.027 &  & -0.3 & 0.029 &  & 0.7 & 0.028 \\
 & $\text{FI}_{m}$ &  & 0.2 & 0.037 &  & 1.1 & 0.059 &  & 4.0 & 0.049 &  & -2.4 & 0.055 &  & -2.5 & 0.042 &  & 0.8 & 0.029 \\ \bottomrule
\end{tabular}%
}
\end{table}

Table \ref{sim2_response} presents the Monte Carlo means of RB and RMSE used to assess the robustness of the methods under a mis-specified response model. The results show that,  across all scenarios, one of three \(\text{PPM}_{\lambda}\) estimators with a correctly specified $\lambda$ remains nearly unbiased, while the other two estimators demonstrate larger biases. However, these biases decrease as \(\rho\) increases from 0.5 to 0.8. This is because \(\text{PPM}_{\lambda}\) only requires the response probability to be an unspecified function of a convex combination of the proxy variable and \(y\). Therefore, the bias caused by response model misspecification diminishes when $y$ is highly correlated with its proxy. 

In contrast, the fractional imputation estimators generally show comparable bias to that of estimates produced by a correctly specified \(\text{PPM}_{\lambda}\) estimator. {Exceptions occur in cases involving Probit or Log-log link functions and the response mechanism $\text{RM}_2$, where the $\text{FI}_m$ estimator shows a relative bias of up to 27.7\%.} Even in these less favorable scenarios, the bias of the \(\text{FI}_m\) estimators remains relatively small compared to that of the \(\text{PPM}_{\lambda}\) estimators under an incorrectly specified $\lambda$. These findings suggest that the proposed FI estimators are likely to be more robust to response model misspecification, particularly given the absence of a definitive method for correctly selecting \(\lambda\) within the PPM framework \citep{little1994class}.  This advantage of the fractional estimators become more pronounced  as the correlation between the proxy variable and the true outcome variable $y$ decreases.

\begin{table}
\caption{\label{sim2_response}
Monte Carlo means of Relative Bias (RB, \%) and Root Mean Square Error (RMSE) of the point estimators for $E(y)$ in cases where the response model is misspecified.}
\centering
\resizebox{\columnwidth}{!}{%
\begin{tabular}{@{}cllrrlrrlrrlrrlrrlrr@{}}
\toprule
 &  &  & \multicolumn{8}{c}{$\rho=0.5$} &  & \multicolumn{8}{c}{$\rho=0.8$} \\ \cmidrule(lr){4-11} \cmidrule(l){13-20} 
 & \multicolumn{1}{c}{} &  & \multicolumn{2}{c}{$\text{RM}_0$} &  & \multicolumn{2}{c}{$\text{RM}_1$} &  & \multicolumn{2}{c}{$\text{RM}_{2}$} &  & \multicolumn{2}{c}{$\text{RM}_0$} &  & \multicolumn{2}{c}{$\text{RM}_1$} &  & \multicolumn{2}{c}{$\text{RM}_{2}$} \\ \cmidrule(lr){4-5} \cmidrule(lr){7-8} \cmidrule(lr){10-11} \cmidrule(lr){13-14} \cmidrule(lr){16-17} \cmidrule(l){19-20} 
Link function & Method &  & RB & RMSE &  & RB & RMSE &  & RB & RMSE &  & RB & RMSE &  & RB & RMSE &  & RB & RMSE \\ \midrule
\multirow{7}{*}{\begin{tabular}[c]{@{}c@{}}Logistic\\ (baseline)\end{tabular}} & CC &  & 17.1 & 0.090 &  & 33.9 & 0.171 &  & 48.2 & 0.242 &  & 27.3 & 0.139 &  & 38.8 & 0.195 &  & 48.3 & 0.242 \\
 & MAR &  & -0.1 & 0.034 &  & 22.3 & 0.115 &  & 38.4 & 0.193 &  & -0.1 & 0.034 &  & 11.0 & 0.061 &  & 20.6 & 0.106 \\
 & $\text{PPM}_{0}$ &  & 0.0 & 0.028 &  & 21.4 & 0.110 &  & 38.3 & 0.193 &  & 0.0 & 0.025 &  & 10.7 & 0.059 &  & 20.3 & 0.104 \\
 & $\text{PPM}_{1}$ &  & -18.6 & 0.098 &  & 5.8 & 0.041 &  & 26.4 & 0.134 &  & -7.6 & 0.046 &  & 2.2 & 0.027 &  & 11.6 & 0.063 \\
 & $\text{PPM}_{\infty}$ &  & -57.8 & 0.294 &  & -29.7 & 0.156 &  & -0.2 & 0.042 &  & -17.3 & 0.091 &  & -9.0 & 0.053 &  & 0.1 & 0.027 \\
 & $\text{FI}_{c}$ &  & -0.2 & 0.051 &  & -0.2 & 0.046 &  & 1.7 & 0.047 &  & -0.1 & 0.035 &  & 0.0 & 0.033 &  & 1.0 & 0.037 \\
 & $\text{FI}_{m}$ &  & -0.1 & 0.044 &  & 0.6 & 0.050 &  & 2.3 & 0.049 &  & -0.1 & 0.035 &  & -0.3 & 0.035 &  & 1.1 & 0.037 \\ \midrule
\multirow{7}{*}{Probit} & CC &  & 22.3 & 0.115 &  & 45.5 & 0.229 &  & 62.8 & 0.314 &  & 35.7 & 0.180 &  & 51.5 & 0.258 &  & 62.9 & 0.315 \\
 & MAR &  & 1.8 & 0.043 &  & 34.2 & 0.173 &  & 52.0 & 0.261 &  & 2.8 & 0.046 &  & 19.3 & 0.102 &  & 31.5 & 0.159 \\
 & $\text{PPM}_{0}$ &  & 0.1 & 0.029 &  & 32.2 & 0.163 &  & 51.7 & 0.259 &  & 0.0 & 0.026 &  & 16.5 & 0.086 &  & 29.4 & 0.148 \\
 & $\text{PPM}_{1}$ &  & -25.7 & 0.134 &  & 12.2 & 0.068 &  & 36.4 & 0.184 &  & -10.6 & 0.060 &  & 4.1 & 0.033 &  & 17.0 & 0.088 \\
 & $\text{PPM}_{\infty}$ &  & -81.9 & 0.416 &  & -38.7 & 0.204 &  & -0.2 & 0.045 &  & -24.5 & 0.126 &  & -12.8 & 0.071 &  & 0.1 & 0.029 \\
 & $\text{FI}_{c}$ &  & 1.1 & 0.081 &  & -2.7 & 0.074 &  & 5.5 & 0.088 &  & 2.5 & 0.047 &  & 2.8 & 0.049 &  & 9.9 & 0.072 \\
 & $\text{FI}_{m}$ &  & 1.5 & 0.070 &  & 4.5 & 0.088 &  & 27.7 & 0.182 &  & 2.5 & 0.046 &  & 2.8 & 0.049 &  & 10.3 & 0.075 \\ \midrule
\multirow{7}{*}{Log-log} & CC &  & 20.0 & 0.104 &  & 39.6 & 0.199 &  & 56.6 & 0.283 &  & 32.1 & 0.162 &  & 45.4 & 0.228 &  & 56.6 & 0.283 \\
 & MAR &  & 5.0 & 0.040 &  & 29.9 & 0.152 &  & 46.5 & 0.234 &  & 8.0 & 0.050 &  & 19.9 & 0.103 &  & 29.4 & 0.149 \\
 & $\text{PPM}_{0}$ &  & 0.0 & 0.029 &  & 27.1 & 0.138 &  & 46.1 & 0.231 &  & 0.0 & 0.026 &  & 13.9 & 0.073 &  & 25.7 & 0.130 \\
 & $\text{PPM}_{1}$ &  & -22.9 & 0.120 &  & 9.2 & 0.054 &  & 32.2 & 0.163 &  & -9.4 & 0.054 &  & 3.2 & 0.030 &  & 14.9 & 0.078 \\
 & $\text{PPM}_{\infty}$ &  & -72.2 & 0.367 &  & -34.5 & 0.182 &  & -0.4 & 0.044 &  & -21.6 & 0.112 &  & -11.1 & 0.063 &  & 0.2 & 0.029 \\
 & $\text{FI}_{c}$ &  & 4.9 & 0.057 &  & 2.4 & 0.048 &  & 9.5 & 0.061 &  & 7.9 & 0.050 &  & 8.2 & 0.050 &  & 13.4 & 0.072 \\
 & $\text{FI}_{m}$ &  & 5.0 & 0.050 &  & 4.6 & 0.059 &  & 22.8 & 0.143 &  & 7.9 & 0.050 &  & 8.2 & 0.050 &  & 13.4 & 0.072 \\ \midrule
\multirow{7}{*}{C.log-log} & CC &  & 21.6 & 0.112 &  & 44.8 & 0.225 &  & 60.9 & 0.305 &  & 34.6 & 0.175 &  & 50.5 & 0.253 &  & 61.1 & 0.306 \\
 & MAR &  & -7.6 & 0.087 &  & 29.2 & 0.149 &  & 49.6 & 0.249 &  & -12.2 & 0.111 &  & 7.7 & 0.070 &  & 24.9 & 0.129 \\
 & $\text{PPM}_{0}$ &  & 0.0 & 0.030 &  & 30.6 & 0.155 &  & 49.6 & 0.249 &  & 0.0 & 0.027 &  & 15.4 & 0.081 &  & 27.5 & 0.139 \\
 & $\text{PPM}_{1}$ &  & -24.5 & 0.128 &  & 10.3 & 0.060 &  & 34.8 & 0.176 &  & -10.1 & 0.057 &  & 3.5 & 0.031 &  & 15.8 & 0.082 \\
 & $\text{PPM}_{\infty}$ &  & -77.2 & 0.392 &  & -39.3 & 0.206 &  & 0.2 & 0.044 &  & -23.1 & 0.119 &  & -12.3 & 0.068 &  & 0.0 & 0.028 \\
 & $\text{FI}_{c}$ &  & -6.5 & 0.110 &  & -7.9 & 0.099 &  & -3.0 & 0.095 &  & -11.9 & 0.113 &  & -13.6 & 0.121 &  & -9.3 & 0.120 \\
 & $\text{FI}_{m}$ &  & -7.0 & 0.099 &  & -6.1 & 0.092 &  & -0.2 & 0.089 &  & -12.0 & 0.111 &  & -13.6 & 0.121 &  & -9.2 & 0.119 \\ \bottomrule
\end{tabular}%
}
\end{table}

{
Additionally, we conduct sensitivity analyses using two artificial datasets to numerically illustrate the approach discussed in Section \ref{Response model selection}. For this purpose, we fix the outcome model and link function to the baseline case (a Normal outcome model and a Logistic link function), and set the response mechanism to the $\text{RM}_1$ type. For each value of $\rho = 0.5$ and $0.8$, we select one Monte Carlo sample from the 1,000 previously generated samples and compare the 95\% confidence intervals derived from $\text{PPM}_{\lambda}$ estimators, the MAR estimator, and FI estimators assuming different forms of $h(\bm{x}; \bm{\alpha})$. Specifically, the working response models for the FI estimators are defined as follows: 
\begin{itemize}
    \item $\text{FI}_1$: $\text{logit}\{ P(\delta=1 \mid \bm{x}, y) \} = \alpha_0 + \beta y$
    \item $\text{FI}_2$: $\text{logit}\{ P(\delta=1 \mid \bm{x}, y) \} = \alpha_0 + \alpha_1x_1 + \beta y$
    \item $\text{FI}_3$: $\text{logit}\{ P(\delta=1 \mid \bm{x}, y) \} = \alpha_0 + \alpha_2x_2 + \beta y$
    \item $\text{FI}_4$: $\text{logit}\{ P(\delta=1 \mid \bm{x}, y) \} = \alpha_0 + \alpha_1x_1 + \alpha_2x_2 + \beta y$.
\end{itemize}
}

\begin{figure}[h]
    \centering
    \caption{95\% confidence intervals for two artificial datasets. The red dashed line represents the true value of $E(y)$.}
    \includegraphics[width=\linewidth]{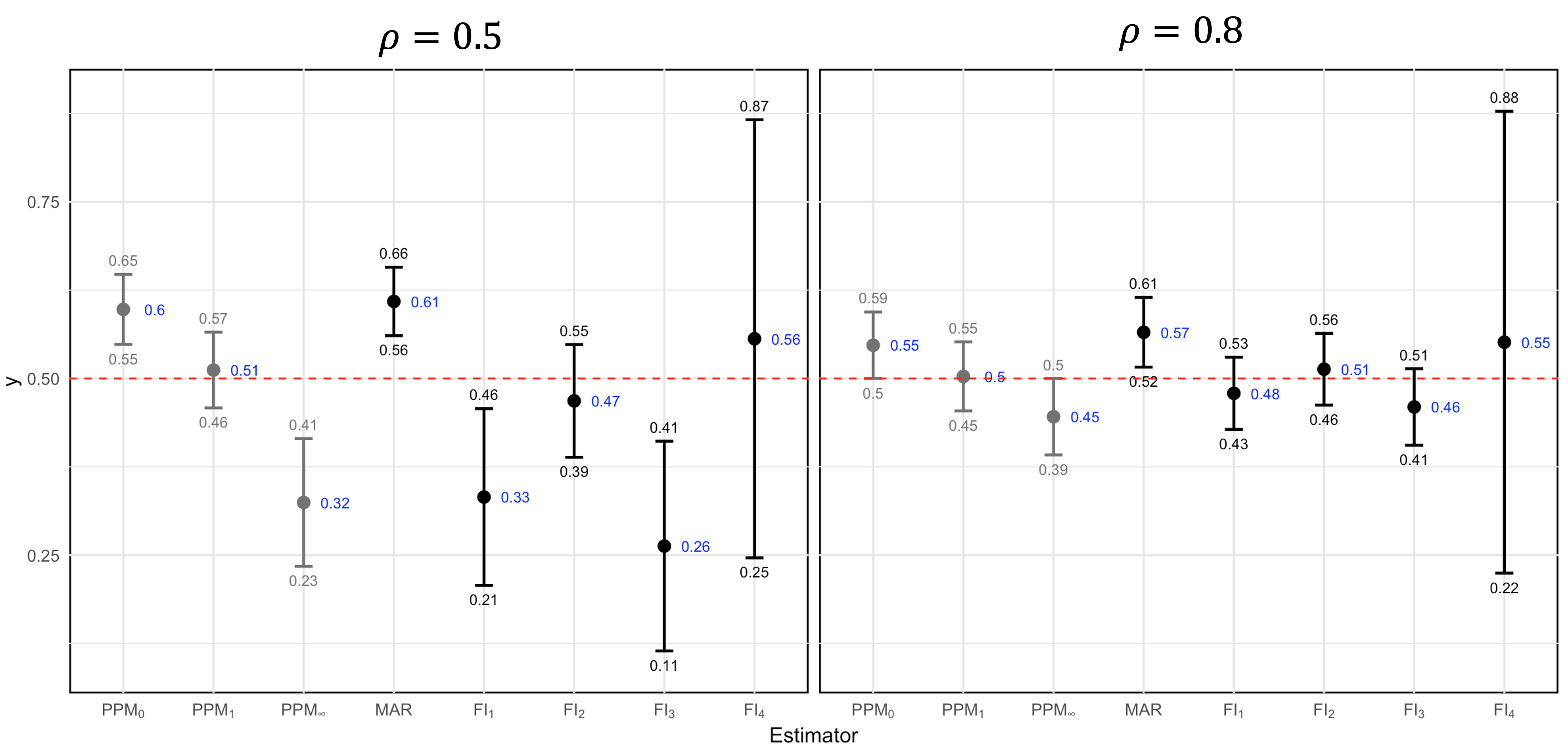}
    \label{Sim_SA}
\end{figure}

{
Figure \ref{Sim_SA} presents the 95\% confidence intervals obtained from each estimator for the two artificial datasets. Notably, the confidence intervals from $\text{FI}_4$, which is based on an unidentifiable model, are unreasonably large compared to those from the other estimators. This pattern is consistently observed across all Monte Carlo samples, as verified by the much larger median confidence interval length for $\text{FI}_4$, shown in Table \ref{median_length}. These results demonstrate that sensitivity analysis can effectively identify unidentifiable models, allowing their exclusion from analysis, as estimates from such models are inconsistent.
}

{
In both artificial datasets, our proposed model selection procedure correctly identifies $\text{FI}_2$, which specifies the true response model. However, when the proxy is weak ($\rho = 0.5$), the confidence intervals exhibit considerable variability depending on the assumed response model. In this case, confidence in the selected model is reduced, as selecting an incorrect model could lead to significant bias. Conversely, when the proxy is strong ($\rho = 0.8$), the confidence intervals for identifiable models are consistent across different response model assumptions. Under such conditions, a single conclusion from our proposed procedure can be made with greater confidence.
}

\begin{table}[h]
\centering
\caption{Median length of 95\% confidence intervals for 1,000 Monte Carlo samples by estimator}
\label{median_length}
\begin{tabular}{@{}clcccccccc@{}}
\toprule
$\rho$ &  & $\text{PPM}_0$ & $\text{PPM}_1$ & $\text{PPM}_{\infty}$ & MAR & $\text{FI}_1$ & $\text{FI}_2$ & $\text{FI}_3$ & $\text{FI}_4$ \\ \midrule
0.5 &  & 0.10 & 0.11 & 0.18 & 0.10 & 0.17 & 0.17 & 0.25 & 0.65 \\
0.8 &  & 0.09 & 0.10 & 0.11 & 0.10 & 0.12 & 0.11 & 0.14 & 0.52 \\ \bottomrule
\end{tabular}
\end{table}

\end{document}